\documentclass[twocolumn]{aastex63} 

\usepackage{xcolor}
\usepackage{amsmath}
\usepackage{latexsym}
\usepackage{amssymb}
\usepackage{enumitem}

\graphicspath{{figures/}}

\DeclareRobustCommand{\disambiguate}[3]{#2~#3}

\usepackage{listings}

\submitjournal{The Astrophysical Journal Letters}
\shorttitle{Self-Supervised Representations of Astronomical Images}
\shortauthors{Hayat et al.}

\begin{document}

\title{Self-Supervised Representation Learning for Astronomical Images}

\author{Md Abul Hayat}
\thanks{Equal contribution first authors.}
\affiliation{University of Arkansas, Fayetteville, AR 72701}
\affiliation{Lawrence Berkeley National Laboratory, Berkeley, CA 94720}
\author{George Stein}
\thanks{Equal contribution first authors.}
\affiliation{Lawrence Berkeley National Laboratory, Berkeley, CA 94720}
\affiliation{Berkeley Center for Cosmological Physics, University of California, Berkeley, CA 94720}
\author{Peter Harrington}
\affiliation{Lawrence Berkeley National Laboratory, Berkeley, CA 94720}
\author{Zarija Luki\'{c}}
\affiliation{Lawrence Berkeley National Laboratory, Berkeley, CA 94720}
\author{Mustafa Mustafa}
\affiliation{Lawrence Berkeley National Laboratory, Berkeley, CA 94720}

\email{mahayat@uark.edu, gstein@berkeley.edu, pharrington@lbl.gov, zarija@lbl.gov, mmustafa@lbl.gov}

\begin{abstract}

Sky surveys are the largest data generators in astronomy, making automated tools for extracting meaningful scientific information an absolute necessity.
We show that, without the need for labels, self-supervised learning recovers representations of sky survey images that are semantically useful for a variety of scientific tasks. These representations can be directly used as features, or fine-tuned, to outperform supervised methods trained only on labeled data. We apply a contrastive learning framework on multi-band galaxy photometry from the Sloan Digital Sky Survey (SDSS), to learn image representations.  We then use them for galaxy morphology classification, and fine-tune them for photometric redshift estimation, using labels from the Galaxy Zoo 2 dataset and SDSS spectroscopy. In both downstream tasks, using the same learned representations, we outperform the supervised state-of-the-art results, and we show that our approach can achieve the accuracy of supervised models while using 2-4 times fewer labels for training. The codes, trained models, and data can be found at \url{https://portal.nersc.gov/project/dasrepo/self-supervised-learning-sdss}.

\end{abstract}


%

\section{Introduction}

Observing and imaging objects in the sky has been the main driver of the scientific discovery process in astronomy,
because doing controlled experiments is not a viable option. The rapid advance of digital sky surveys in the 1990s, spearheaded by SDSS~\citep{Gunn1998, Gunn2006}, has rendered obsolete the approach of manual inspection of images by an expert. Instead, computational analysis methods are constantly being developed and applied~\citep{Ivezic2019}.  Additionally, ``citizen science'' like the Galaxy Zoo\footnote{\url{https://www.zooniverse.org/projects/zookeeper/galaxy-zoo/}} project~\citep[GZ,][]{Lintott2008} plays an important role for tasks which are too complex to describe algorithmically, yet are heuristically quite comprehensible to humans, such as classification of galaxies based on their morphological types~\citep{Lintott2013}. In recent years, machine learning methods have proven particularly useful for both classification and regression tasks (see \citealp{ml-in-cosmo} for a comprehensive list), but the majority of published works rely on the quantity and quality of (manually assigned) image labels.

Serendipitous discovery of an ionization echo from a recently faded quasar \citep{Lintott2009}, and the cumbersome search for similar systems that followed \citep{Keel2012}, showcases other big data challenges. It demonstrates the need for methods which allow for the discovery of truly unusual and previously unseen objects, and also the need to perform semantic (or feature) similarity searches on images in situations when the number of labels is as low as one. In the near future, incoming sky surveys such as the Vera Rubin Observatory \citep{LSST2019}, Euclid \citep{Euclid2011}, Nancy Grace Roman Space Telescope \citep{Spergel2013}, or the Square Kilometer Array\footnote{https://www.skatelescope.org/} will open yet another research epoch, where datasets are of sizes which completely overwhelm even the most ambitious citizen science concepts.  It is fair to say that the vast majority of images from these observatories will never be seen by a human eye. Thus, the capability to organize images without labels and programmatically search for semantic similarity or for interesting outliers will be essential to maximize the scientific output of these missions.

This capability is heavily dependent on image \emph{representations} — low-dimensional mappings of images which preserve their inherent information. Finding good representations is crucial to scientific \emph{downstream} tasks such as clustering and classification of images, but is often elusive, partly due to the difficulty of gathering enough high-quality labels. Unsupervised machine learning methods aim to learn semantically meaningful representations of the data without relying on any labels \citep[see, e.g.][]{Alloghani2020}. Many such methods have already been applied to studies of galaxy morphology \citep{Hocking2018, Martin2020, Cheng2020, Spindler2020}, identification of strong lenses \citep{Cheng2020b}, and anomaly detection \citep{Xiong2018, MargalefBentabol2020}. Unfortunately, across most computer vision applications, the utility of unsupervised representations for downstream tasks has historically lagged behind that of the representations coming from supervised training \citep{Caron2018_unsup}.

However, very recent progress in self-supervised learning has now closed the gap with supervised learning in computer vision \citep{he2020momentum, chen2020simple, chen2020improved, chen2020big}. Self-supervised methods learn representations by training models to solve contrived tasks (e.g., filling in empty regions of data samples, or identifying different versions of the same object as a pair) where the labels are generated algorithmically from an unlabeled dataset. The aim is to design models and tasks that yield semantically meaningful representations which are useful for a variety of downstream tasks, and can be directly used or fine-tuned for these applications. Self-supervised pre-training is vital to state-of-the-art natural language models~\citep{BERT, GPT, GoogleSearchBERT}; now that this method has undoubtedly crossed over into the computer vision domain, it has exciting prospects for broad scientific use.

In this paper, we demonstrate that self-supervised learning indeed has great utility for large astronomical surveys, using $\sim 1.2$ million SDSS \emph{ugriz} galaxy images with $64\times64$ pixels as a proof of concept dataset (full details of data acquisition and selection are given in Appendix~\ref{app:data}). In section \ref{sec:method}, we review the method of contrastive self-supervised learning and propose data augmentations that induce good representations for sky survey images. This approach allows us to build powerful representations which we showcase in section \ref{sec:visualization}. In sections~\ref{sec:morphology} and \ref{sec:redshifts} we use the self-supervised representations to quickly outperform supervised learning at two very common downstream tasks: morphological classification and inference of photometric redshifts, respectively.

\section{Method}
\label{sec:method}

\begin{figure*}[t]
\centering
\includegraphics[width=0.9\textwidth]{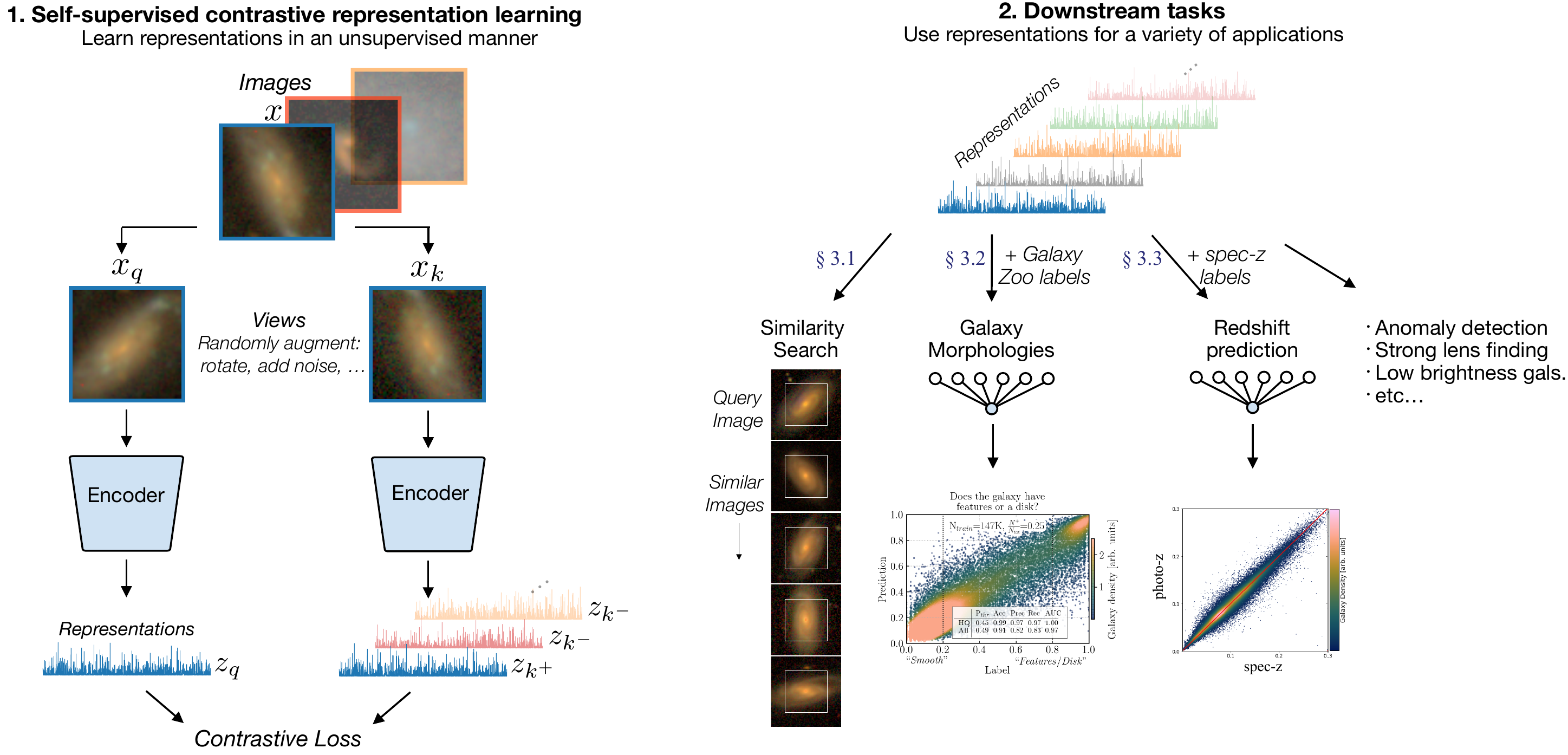}
\caption{(Left) A schematic of the contrastive self-supervised framework. (Right) Examples of downstream tasks that can be implemented on the learned representations.}
\label{fig:network}
\end{figure*}

Recent self-supervised works~\citep{NEURIPS2019_ddf35421,Misra2019, he2020momentum, chen2020simple, chen2020improved, chen2020big} use contrastive losses~\citep{ContrastiveLoss} to minimize the distance between different \emph{views} (augmentations) of the same image in a learned representation space, while maximizing the distance between the representations of different images. The randomized augmentations producing these views should be semantic-preserving transformations of the input images, and the goal is to make the final representation invariant to these transformations \citep{Tian2020, Xiao2020}. This key design choice is application-dependent and requires prior knowledge.  For example, in a galaxy survey, changing colors of galaxies could be detrimental for the downstream task of inferring photometric redshifts, even though color augmentation may be useful when classifying cats and dogs. For a base set of image augmentations that would be useful to the vast majority of downstream applications in sky surveys we propose the following:
 \begin{itemize}[leftmargin=*]
    \itemsep0em 
    \item \textbf{Galactic extinction}.  We want features to be invariant to the galactic latitude and object's position on the celestial sphere.  To model the effects of foreground galactic dust, we introduce artificial reddening by sampling a $E(B-V)$ reddening value from $\mathcal{U}(0,0.5)$ and applying the corresponding per-channel extinction according to the photometric calibration from \cite{schlafly2011measuring}.
    \item \textbf{Point Spread Function} (PSF). Due to a variety of factors over the time span of a galaxy survey, images do not have a consistent PSF. To be invariant to this we experiment with a PSF augmentation, modeled as wavelength-dependent Gaussian smoothing with a standard deviation in $r$-band drawn from $\mathcal{N}(0, 0.13'')$ and scaled appropriately to the other channels using $\lambda^{-0.3}$ \citep{sdsspsf}.    
    \item \textbf{Rotation}.  To be invariant to the apparent orientation of each galaxy, we sample the angle of random rotation of each image from $\mathcal{U}(0,2\pi)$. 
    \item \textbf{Random jitter \& crop}.  We also desire invariance to the image centering. Thus, two integers are sampled from $\mathcal{U}(-7,7)$ to move (jitter) the center of the image (of size 107$^2$) along each respective axis, then the jittered image is center-cropped to size 64$^2$.
    \item \textbf{Gaussian noise}.  Finally, to be invariant to the instrumental noise, we sample a scalar from $\mathcal{U}(1,3)$ and multiply it with the aggregate median absolute deviation (MAD) of each channel (pre-computed over all training examples) to get a fixed per-channel noise scale $\gamma_c$. Then, we introduce Gaussian noise sampled from $\mathcal{N}(0,\gamma_c)$ for each color channel.
\end{itemize}

The relative importance of these augmentations for producing good representations depends on both the dataset and the implementation of each augmentation. We evaluate representation quality by fine-tuning our representations for the task of redshift estimation under limited data labels (see Appendix~\ref{app:ablation} for details), finding Gaussian noise to be our strongest data augmentation and PSF the weakest, likely because pooling layers in our convolutional neural networks (CNNs) are robust to small-scale smearing. Best quality is achieved when we apply all augmentations except PSF. Note that these findings will not necessarily generalize to other surveys with different resolutions, signal-to-noise ratios, or target objects. This base set of image augmentations was chosen to remain as task-agnostic as possible, and additional augmentations could be added to target specific applications. For example, in tasks where the angular extent of a galaxy is irrelevant, an augmentation to change the apparent galaxy size (via image rescaling/interpolation) would be useful.

A schematic of the self-supervised pre-training framework used is shown in Figure~\ref{fig:network} (Left). Applying our augmentations to samples $\bf{x}$, we get a pair of views that are denoted ``positive’’ (${\bf {x}_q}$, ${\bf {x}_{k^+}}$) when the two come from different transformations of the same image, and ``negative’’ (${\bf {x}_q}$, ${\bf {x}_{k^-}}$) otherwise. For each of the views, an encoder network extracts a $2048$ dimensional representation ${\bf z} = \text{encoder}(\bf x)$, and is trained to make positive pairs have similar representations while making negative pairs have dissimilar representations via a contrastive loss function:

\begin{align}
\label{eq:loss}
 \begin{split}
    &L_{q,k^+,\{k^-\}} = \\
    -\log &\bigg( \frac{\exp( \text{sim}(\bf{z}_q , \bf{z}_{k^+}))}{\exp( \text{sim}(\bf{z}_q , \bf{z}_{k^+})) + \sum_{k^-} \exp( \text{sim} (\bf{z}_q ,\bf{z}_{k^-}))} \bigg), 
 \end{split}
\end{align}
where $\text{sim}(\bf{a},\bf{b}) = \bf{a} \cdot \bf{b}/ (\tau\| \bf{a} \| \, \| \bf{b} \| )$ is the cosine similarity measure between vectors $\bf{a}$ and $\bf{b}$, normalized by a tunable ``temperature'' hyper-parameter $\tau$. This loss \citep[InfoNCE,][]{oord2018representation} is minimized when positive pairs have high similarity, while negative pairs have low similarity. We have closely followed \cite{chen2020improved} in our self-supervised learning setup, and more implementation details are given in Appendix~\ref{app:architecture}.

\section{Results}
\label{sec:results}


We first visualize how the model has organized the image representation space, and explore how morphological characteristics from the Galaxy Zoo 2 project \citep[GZ2,][]{GZ2} and spectroscopic redshifts from SDSS map onto this representation space. Then, using the labels from these two sources, we evaluate the utility of our self-supervised representations in actually performing the downstream tasks of morphology classification and photometric redshift estimation.

\subsection{Self-supervised learning visualization}
\label{sec:visualization}
\begin{figure*}[t]
\centering
\includegraphics[width=1.0\textwidth]{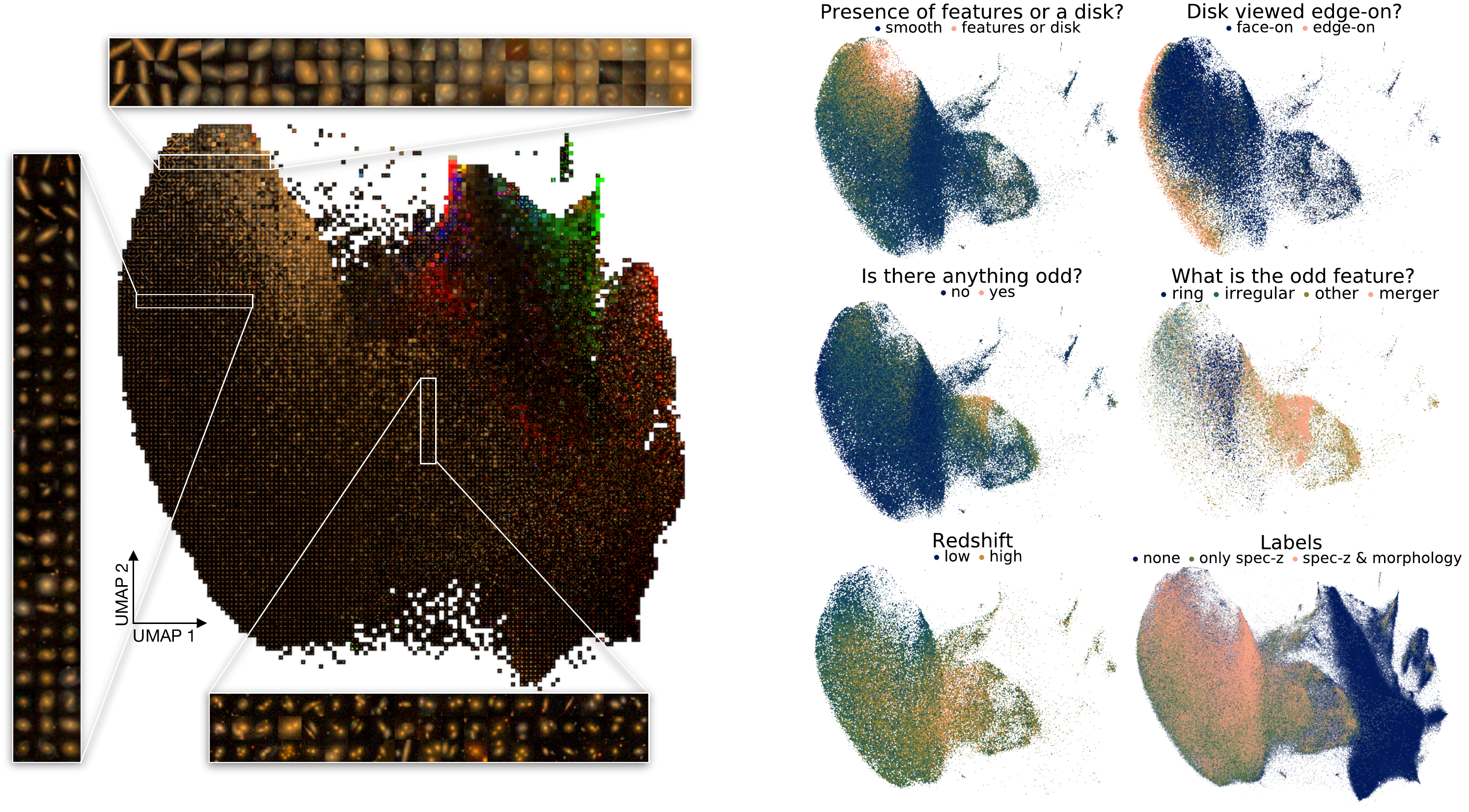}
\caption{Visualizing the two dimensional UMAP projection of the self-supervised representations. The left panel shows randomly sampled representative images at each point in the space, while the right colors the space using answers to morphological classification questions from Galaxy Zoo 2, SDSS spectroscopic redshifts, or by labels.} 
\label{fig:umap}
\end{figure*}

To visualize the information contained in the self-supervised representations we use Uniform Manifold Approximation and Projection \citep[UMAP,][]{UMAP} to reduce the 2048 dimensional representations to a more manageable 2, while preserving structure information on both local and global scales. We want to emphasize that although UMAP can produce meaningful clusters when trained directly on image data, we are using it here only for visualization purposes of the representation space. The fact that the morphological classification tasks described in the next section can achieve a high performance through only a linear transformation of the representations, with no fine-tuning, means that the galaxies are organized in a semantically meaningful way in the representation space.

In Figure~\ref{fig:umap} we investigate this 2D projection. The left panel was created by binning the space into 128$\times$128 cells, randomly selecting a sample that resides within each cell, and plotting its corresponding rgb mapped galaxy image at that location\footnote{rgb images are obtained by `luptonizing' \citep{luptonize} the \textit{gri} photometric bands}. Around the edges we show zoom-ins to a variety of hand-selected areas, in which it is clear that images are grouped by their visual similarity (e.g., spiral or not, edge-on or not, etc).

The following six panels color each point using the redshift and morphology labels, and confirm that clustering is not only along visual characteristics. Distinct clusters as a function of morphological type and redshift are immediately apparent, to the level where decision boundaries for a number of GZ2 questions can be drawn by eye. Morphological labels are uncertain, so we illustrate them as continuous colors representing the fraction of votes for one answer over the other. Interestingly, we see that a large number of unlabeled samples are separated from any that have either redshift or morphology labels, but as we show below, using them for self-supervised learning still proves beneficial for the downstream tasks. 

Appendix~\ref{app:umap} further examines this 2D space in context of galaxy size and magnitude, and shows the advantage over the equivalent UMAP representations derived instead directly from the pixel space. We also demonstrate how a sample of galaxies under simple augmentations move drastically through this plane when the UMAP is derived from the images, but remain stationary when using the self-supervised representations.

When displayed through an interactive data portal \citep[e.g.][]{galaxyportal}, such visualizations built upon self-supervised representations can be invaluable to the broader astronomical community.

\begin{figure*}[t]
\centering
\includegraphics[trim={0cm 0 0 0},clip, width=0.95\textwidth]{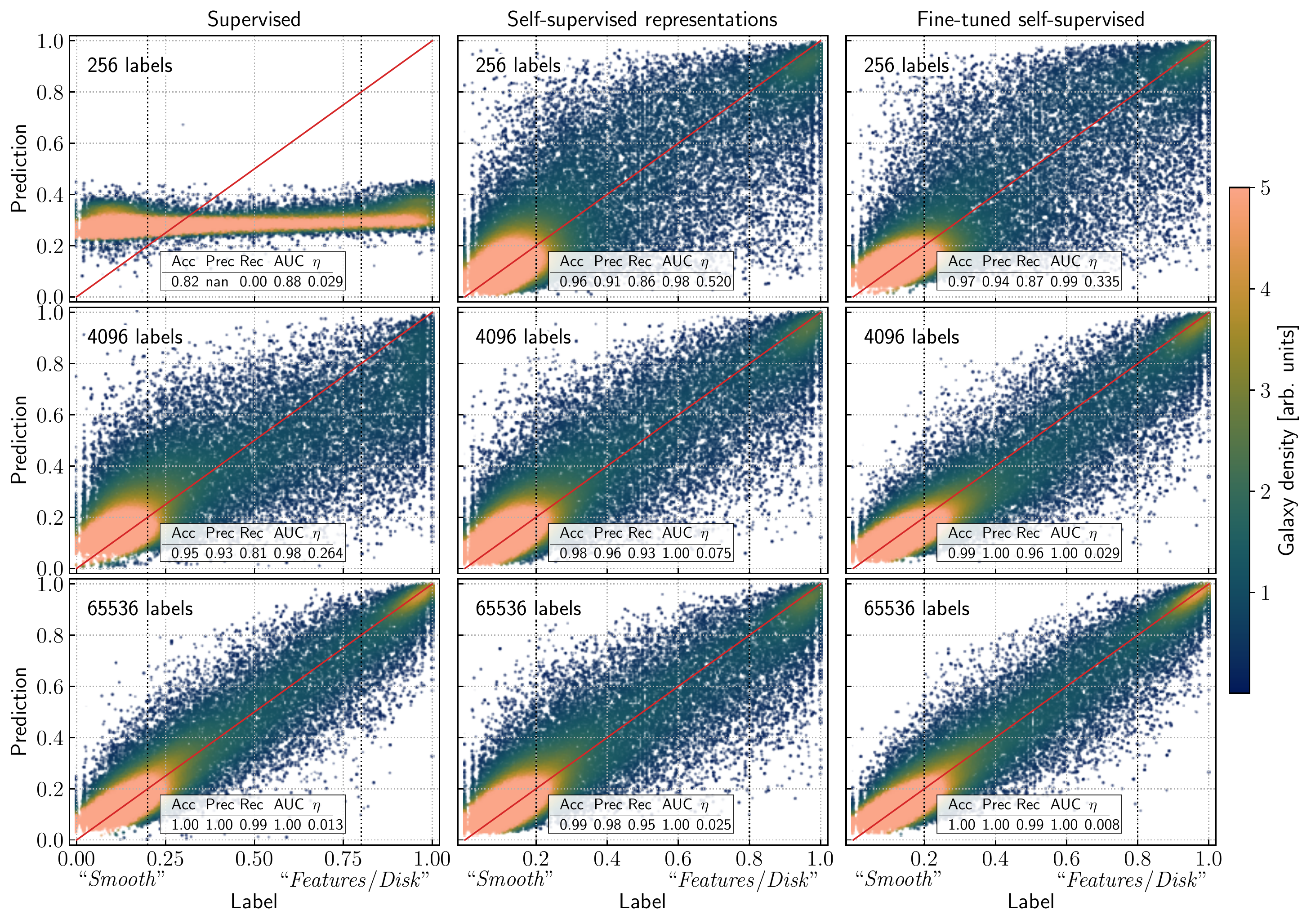}
\caption{Predicted labels compared to crowd-sourced answers for the first GZ2 question: ``Is the galaxy simply smooth and rounded, with no sign of a disk?''. The three columns show the classification performance in a supervised setting (left), a linear classifier {\textit{directly on the self-supervised representations}} (center), and when fine-tuning the self-supervised encoder for a few epochs (right). We report the accuracy, precision, recall, AUC, and outlier percentage $\eta$ for ``high quality'' labels (those with $P > 0.80$ or $P < 0.2$). The size and opacity of the points is proportional to the number of crowd-sourced labels they received.}
\label{fig:morphology}
\end{figure*}

\subsection{Galaxy morphology classification}
\label{sec:morphology}
Morphological classification of galaxies into subclasses based on the presence of visual characteristics such as spiral arms, central bars, or odd features, is key in order to study galaxy formation and evolution. The Galaxy Zoo project \citep{Lintott2008} has been fundamental in this endeavour by crowd-sourcing morphological classifications for a significant number of galaxies. GZ2 \citep{GZ2}, the successor to GZ, is focused on more fine-grained features, and in total achieved morphological classifications of 304,122 SDSS galaxies. Shown most prominently by the winners of the ``Galaxy Challenge'' \citep{galaxy_challenge} and numerous subsequent works since \citep{DS18, DS_transferlearning, K19, W20, Spindler2020, des_morphology}, CNNs excel at this task. 

Here, by treating each question as a separate binary classification task, we predict answers to the subset of GZ2 questions that are most commonly undertaken by ML methods. We train three separate classifiers. The first is a CNN trained from scratch in a supervised setting with the same architecture of the encoder, the second is a linear classifier applied {\textit{directly on the self-supervised representations}}, and for the third we fine-tune the self-supervised encoder for a few epochs. We note that the linear classifier requires only $\sim 0.5-10$ seconds on a GPU to train depending on the number of training samples used, while the fully supervised training takes up to 2 hours on 8 GPUs.

Figure~\ref{fig:morphology} demonstrates the quality of the morphological predictions for the first GZ2 question, and shows the predicted label against the true label for the three classifiers as a function of the number of labels used for training. We quantify the accuracy, precision, recall, area under the receiver operator characteristic curve (AUC), and the outlier percentage $\eta$ in the inlaid tables (see Appendix~\ref{app:morphology} for definitions: 1.00 is the ideal value for the first four, and 0 is ideal for $\eta$). Our results should be viewed most closely in relation to \citet[DS+18]{DS18} and \citet[W+20]{W20}, as both used SDSS images and focused on GZ2 questions. The  performance metrics shown are calculated on ``high quality'' labels with $P > 0.80$ or $P < 0.2$, although the networks were trained using all labels for a given GZ2 question with at least 5 votes regardless of the vote fraction for one answer over the other.

\begin{figure*}[t]
\centering
\includegraphics[width=\linewidth]{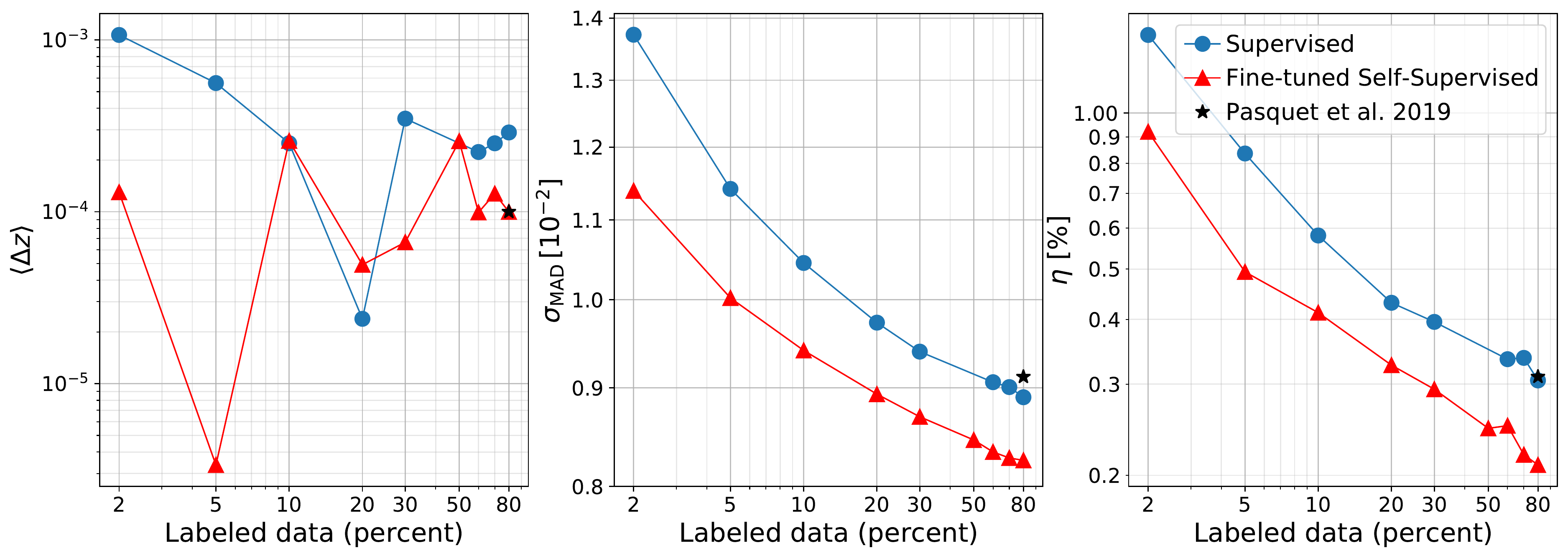}
\caption{The prediction  bias $\langle \Delta z \rangle$, dispersion $\sigma_{\rm MAD}$, and outlier percentage $\eta$ of photo-$z$ estimates on test data, from our fine-tuned representations compared to reference fully-supervised networks. Trained on increasing fractions of the spec-$z$ dataset, the fine-tuned self-supervised models perform best.
}
\label{fig:power_law}
\end{figure*}

We find that both the linear classifier from the representations and the fine-tuned self-supervised model far outperform the supervised network when training with a limited number of labels, and are both able to achieve accurate classifications even in the regime where the supervised network fails to converge. Comparing the supervised network to the fine-tuned self-supervised for this GZ question, we find that roughly a factor of 16 more labels are required in the supervised setting to achieve the same performance as fine-tuned. When increasing to 65k labels we find that the supervised and fine-tuned networks are approaching optimal classification performance on this dataset given the high level of label uncertainty introduced by ambiguous class boundaries and crowd-sourced labeling. We note that an exact quantitative comparison of our performance metrics to DS+18 and W+20 are not possible due to a lack of consistency in data sets. In Appendix~\ref{app:morphology} we present additional discussion, methods, and morphological results on other GZ2 questions.

These results demonstrate that the self-supervised representations are extremely valuable for morphological classification. (1) They are essential to make accurate predictions when restricted by the number of available labels; (2) they improve accuracy metrics beyond what was achieved by pure supervised learning in the non-optimal classification performance regime; (3) they provide avenues to investigate and isolate imaging artifacts and anomalies as shown in the Appendix \ref{ap:galaxy_morphology}; (4) they can be used in pipelines to speed up crowd-sourced classification tasks: determine the next galaxy to be classified, perform a very computationally inexpensive similarity search\footnote{we use faiss: \url{https://github.com/facebookresearch/faiss}} to find $N$ other similar galaxies, and classify them all at once; and (5) they can reduce the barrier to entry when analysing survey data by achieving high classification accuracy through linear classifications on the representations which requires minimal machine learning experience and compute resources.

\subsection{Photometric Redshift Estimation} \label{sec:redshifts}

\begin{figure*}[t]
\centering
\includegraphics[width=0.8\linewidth]{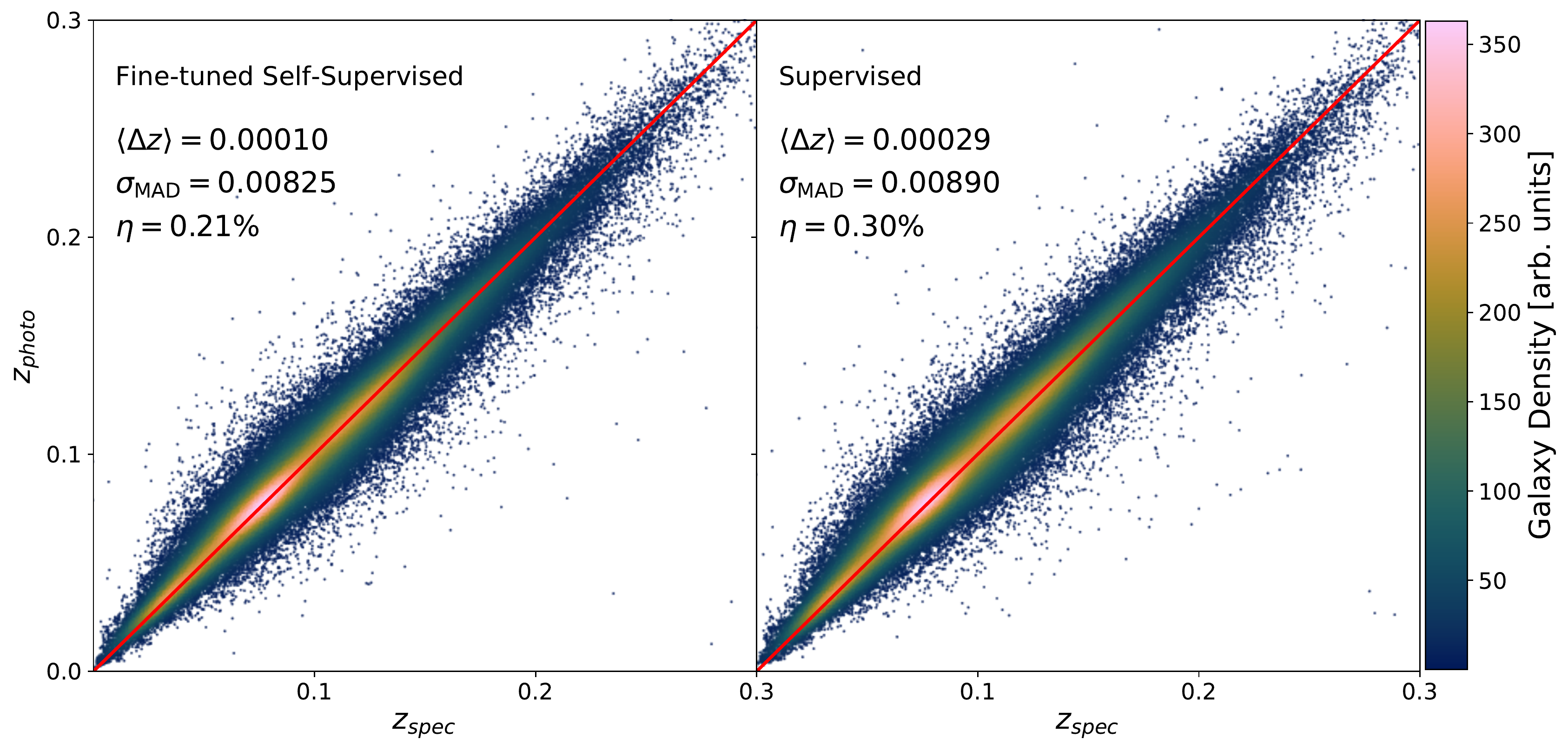}
\caption{The photo-$z$ estimates of our best model on test data, arising from fine-tuned self-supervised model (left), compared to the equivalent supervised network (right).}
\label{fig:photoz_vs_specz}
\end{figure*}

Determining the redshifts, and hence distances, to the billions of galaxies imaged in cosmological surveys is crucial to studying large-scale-structure, but taking high-precision spectroscopic redshift (spec-$z$) measurements for each galaxy is infeasible. Thus, a task of great importance to sky surveys is photometric redshift (photo-$z$) estimation \citep[for a recent review, see][]{salvato2019many}. 
Photo-$z$ models take multi-band galaxy photometry and use template fitting \citep{Loh1986}, machine learning \citep{Connolly1995}, or hybrid methods to produce estimates for the galaxy’s redshift $z$ and its associated probability density function. Traditionally, such models relied on ``hand-crafted’’ features extracted from the observations, but recent work has also successfully applied CNNs directly to the images themselves \citep{HoyleCNN2016, DIsantoCNN2018, pasquet2019photometric}. While promising, these supervised methods are inherently constrained by the limited size (and galaxy features, such as relative size or brightness) of the training dataset. Therefore, self-supervised representations derived from a larger body of unlabeled samples are a promising venue for improving accuracy and robustness.

For easier comparison against established baselines on SDSS data, we closely follow the setup of  \cite{pasquet2019photometric}, whose CNN achieved significantly lower dispersion than previous image-based ML models. Their network is trained as a classifier over a discrete set of 180 redshift bins spanning $0\leq z \leq 0.4$, where the photo-$z$ estimate $z_p$ is computed as the expectation $\mathbb{E}(z)$ over the probabilities predicted in each bin. We adapt this design into our ResNet50 model, and establish our own supervised baseline (with identical architecture as our self-supervised model) by training on the available spec-$z$ labels. We use the following standard metrics to evaluate the accuracy of photo-$z$ estimates:

\begin{itemize}
\item The prediction residual $\Delta z = (z_{p}-z_{s})/(1+z_{s})$, where $z_p$ and $z_s$ correspond to the photometric and spectroscopic redshifts, respectively.
\item The dispersion or MAD deviation, $\sigma_{\rm MAD}= 1.4826 \times \rm{MAD}(\Delta z)$,  where MAD $ = \mathrm{median} ( |\Delta z - \mathrm{median}(\Delta z )|)$.
\item $\eta$, the percent of ``catastrophic'' outliers with $|\Delta z| > 0.05$.
\end{itemize}

Results of our supervised training study are shown in the three left panels of Figure~\ref{fig:power_law}. As shown, we also test the improvement in the model accuracy as we increase the training dataset size. Similar to the results of~\cite{pasquet2019photometric}, the prediction bias $\langle \Delta z \rangle$ (a noise-dominated metric) is negligibly small, and our supervised baseline model matches the accuracy of~\cite{pasquet2019photometric} in $\sigma_{\rm MAD}$ and $\eta$.

We then move to evaluating the utility of our self-supervised representations for photo-$z$ estimation. To fine-tune them on the spec-$z$ labels, we train a linear classifier on the representations while allowing the encoder weights to train with a $10\times$ smaller learning rate than that of the classifier (more fine-tuning details are in Appendix \ref{app:architecture}). Results are shown in Figure~\ref{fig:power_law}.

At all fractions of training data used, the fine-tuned self-supervised representations achieve superior performance in $\sigma_{\rm MAD}$ and $\eta$ compared to the equivalent supervised network. Impressively, the accuracy gained from pre-training on unlabeled data is equivalent to supervised training on $2-4\times$ more spec-$z$ labels, with no modifications in architecture size or complexity. Consequently, our model, fine-tuned on the full training dataset, achieves a new state-of-the-art accuracy for CNN-based photo-$z$ prediction on SDSS galaxies, as shown in Figure~\ref{fig:photoz_vs_specz}. This is an exciting result, as it suggests any existing supervised network developed for similar tasks on this type of data could get an immediate performance benefit from a self-supervised pre-training stage.

\section{Discussion and Conclusions}
In this letter we have demonstrated that self-supervised representation learning on unlabeled data yields notable performance gains over supervised learning for multiple tasks. These performance gains are achieved even when the self-supervised model is limited to have the same size as the baseline CNNs in downstream tasks. However, results from ML literature show that the best performance (i.e., the best representation quality) is achieved when self-supervised models are much larger \citep{radford2019language, chen2020big}. Thus,  the possibility of training a large self-supervised model on massive photometry databases and “serving” the model for usage by the larger community, much like the operation of existing state-of-the-art language models \citep{bert2018, radford2019language}, is an exciting new direction for ML applications in sky-surveys.

A major issue with all machine learning studies on labeled sky survey data is not just the limited size of the training data set, but also the selection bias imposed by gathering labels. For example, due to the series of flux and quality cuts applied when selecting spectroscopic targets \citep{Strauss2002_SDSS}, galaxies with spec-$z$ labels have a distinct bias towards nearby bright objects with low galactic extinction. Thus, galaxies selected for spectroscopic measurement are not representative of all those with photometry, as can clearly be seen in the representation space of Figure~\ref{fig:umap}. Although ML methods can train on this labeled data and achieve a good test accuracy within this subset of galaxies, there are few robustness guarantees for photo-$z$ estimation beyond the labeled distribution of galaxies. It has been shown, recently, that self-supervised pre-training \citep{Hendrycks2019pretrain, Hendrycks2019selfsup} improves model robustness and uncertainty quantification, and that self-supervised models outperform their supervised counterparts in out-of-distribution detection on difficult outliers. This means self-supervised models have excellent prospects for mitigating distributional differences between (labeled) training and (unlabeled) inference data in sky surveys.
We believe that self-supervised representation learning opens the door to leveraging the vast amounts of unlabeled, existing and future, sky survey data, promising a new era for ML applications in precision and discovery astrophysics.

\acknowledgments
Authors would like to thank Fran\c{c}ois Lanusse, Peter Melchoir, Evan Racah, and Edward Schlafly for helpful discussions. This research used resources of the National Energy Research Scientific Computing Center (NERSC), a U.S. Department of Energy Office of Science User Facility located at Lawrence Berkeley National Laboratory, operated under Contract No. DE-AC02-05CH11231. Md.H.'s work was supported by the NERSC's summer internship program. G.S.~and Z.L.~were partially supported by the DOE's Office of Advanced Scientific Computing Research and Office of High Energy Physics through the Scientific Discovery through Advanced Computing (SciDAC) program.

\bibliographystyle{aasjournal}
\DeclareRobustCommand{\disambiguate}[3]{#3}
\bibliography{ref_apjl}{}

\appendix
\section{Dataset details}
\label{app:data}

Our database of galaxies is assembled from Data Release 12 (DR12; \citealt{alam2015eleventh}) of the SDSS. To pull samples with spectroscopic redshift labels, we follow the process of \cite{pasquet2019photometric} in pulling from the Main Galaxy Sample to enable direct comparison to their results. Their SQL query filters for objects classified as \texttt{`GALAXY'} with dereddened petrosian magnitudes $r \leq 17.8$ and spectroscopic redshifts $z \leq 0.4$. For us, the query returns 547,224 objects, and after removing some duplicates, we are left with  517,190 to use as labeled training examples. When fine-tuning our image representations for the photo-$z$ estimation task, we use  400,000 images for training and 103,000 as validation dataset.

\begin{figure}
    \centering
    \includegraphics[width=0.49\textwidth]{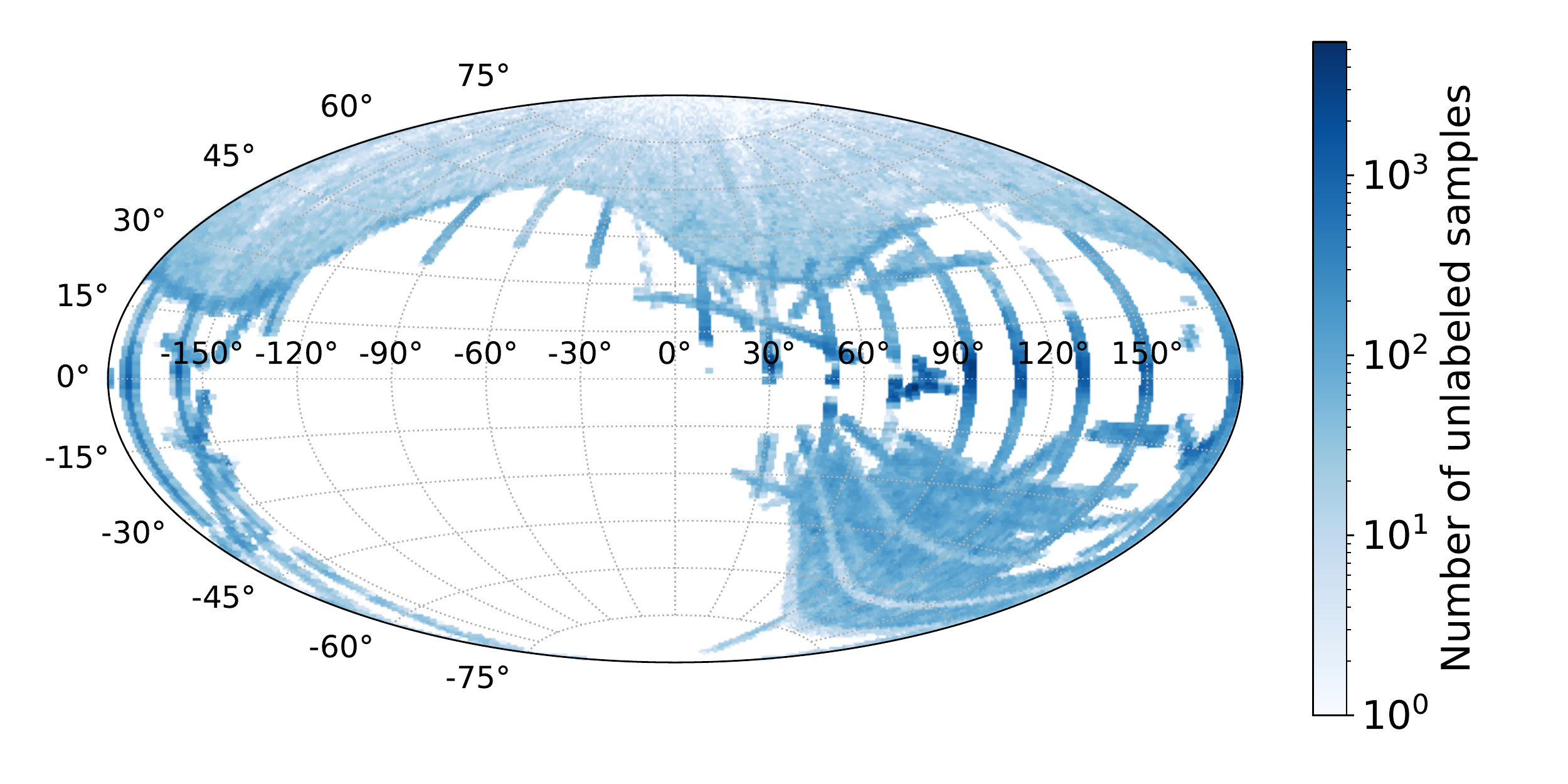}
    \includegraphics[width=0.49\textwidth]{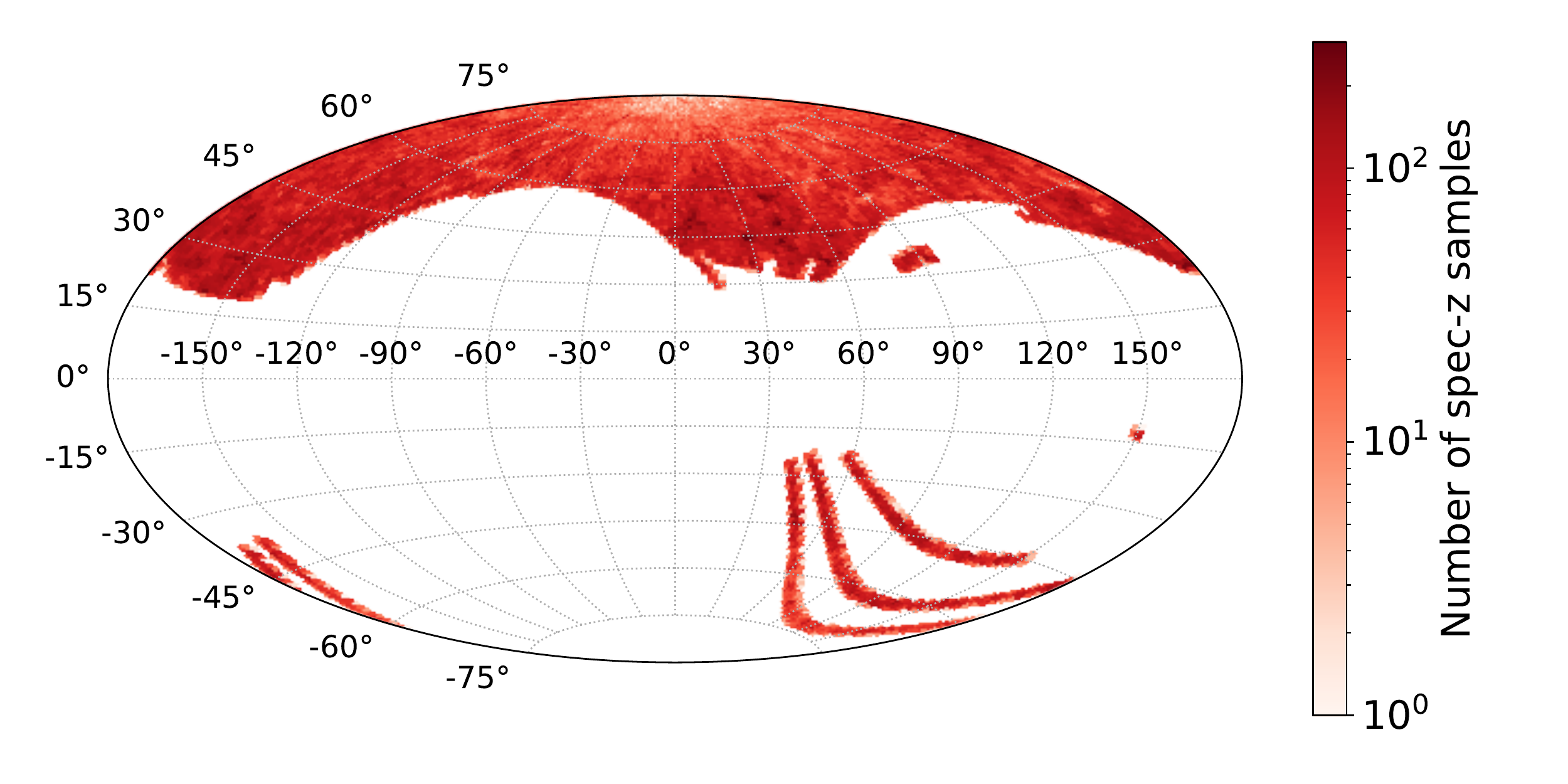}
    \caption{The spatial distribution of the full unlabeled (left) galaxy image dataset, and the spec-z labeled (right) dataset.}
    \label{fig:skymap}
\end{figure}

To build our larger set of galaxies with no spectroscopic labels, we filter for objects with dereddened petrosian magnitudes $r \leq 17.8$, on the \texttt{`PhotoObjAll'} full photometric catalog of the SDSS. In the resulting set of galaxies, we remove duplicates which were already included in our spectroscopic sample, and exclude samples with an estimated photometric redshift (as estimated by \cite{Becketal16photozbaseline}) $z_{phot} > 0.8$. This eliminates objects which are too distant compared to the spectroscopic sample, but decreases the possibility that we are unnecessarily excluding samples whose true redshift is less than 0.4 (the cutoff for our spectroscopic sample) due to incorrect photo-$z$ estimates. After imposing these cuts, we were able to successfully pull 845,254 unlabeled images. The spatial distributions of our labeled and unlabeled galaxy datasets are shown in Figure \ref{fig:skymap}

SDSS photometric images contain data in 5 passbands ($ugriz$), and come background-subtracted but are not de-reddened to account for galactic extinction. To pull images for our datasets, we use the Montage\footnote{\url{http://montage.ipac.caltech.edu/}} tool to query the imagery catalog in SDSS Data Release 9 (DR9), based on the tabulated equatorial coordinates for each object in our dataset. For each set of object coordinates, we sample a patch of sky of size $(0.012^\circ)^2$, centered on the object, and project onto a 2D image with $107^2$ pixels (this ensures the resulting pixel scale is as close as possible to the native pixel scale in the SDSS, 0.396 arcsec). In each image, we store the $u$, $g$, $r$, $i$, and $z$ passbands as 5 color channels.

The Montage pipeline uses Variable-Pixel Linear Reconstruction \citep{FruchterHook2002VLPR} during the projection process to appropriately transform source input pixel values into the output pixel space. With this setup, a few samples were returned containing corrupted values at the edges of the image, so we crop all images to $107^2$ pixels to eliminate such issues. Note that during training of the self-supervised model, we impose random rotations and random jitter to each image before cropping out the central portion as a data augmentation, so while our photometric images contain 107 pixels per side, the CNNs in this work only view samples of size $64^2$ pixels. This input size of CNN is consistent with the photo-$z$ CNN model of \cite{pasquet2019photometric}.

\section{Similarity Search and UMAP}
\label{app:umap}

Section~\ref{sec:visualization} visualized the information contained within the self-supervised representations derived from the SDSS dataset in context of a subset of the available labels. We found that the representations were organized to a high-degree by the semantic information contained within the images. Here we further show that the UMAP performed on the representations surpasses the utility of a UMAP performed directly on the images, and demonstrate how the self-supervised representations allow for simple and effective similarity searches.

\subsection{Similarity search}  
\label{sec:similarity}

\begin{figure}[t]
  \centering
    \includegraphics[width=0.4\textwidth, 
    trim={0.25cm 0.25cm 0.25cm 0.25cm}, clip]{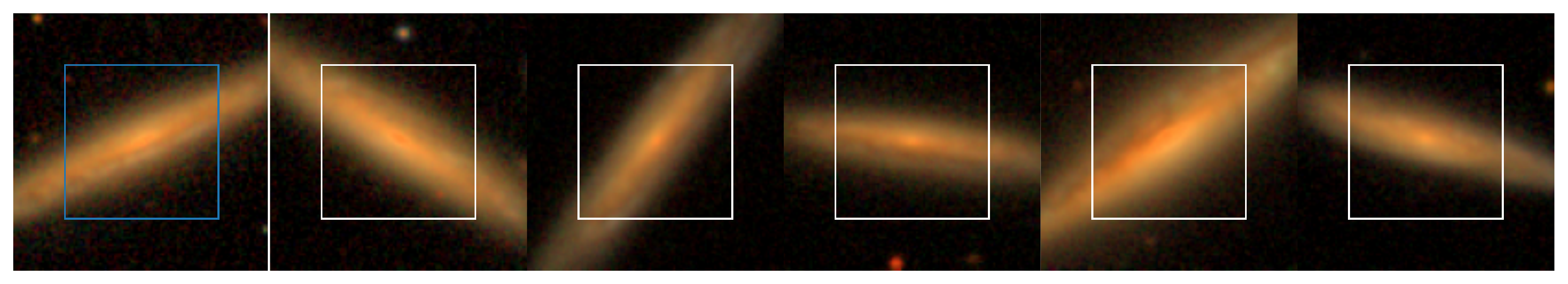}
    \includegraphics[width=0.4\textwidth, 
    trim={0.25cm 0.25cm 0.25cm 0.25cm}, clip]{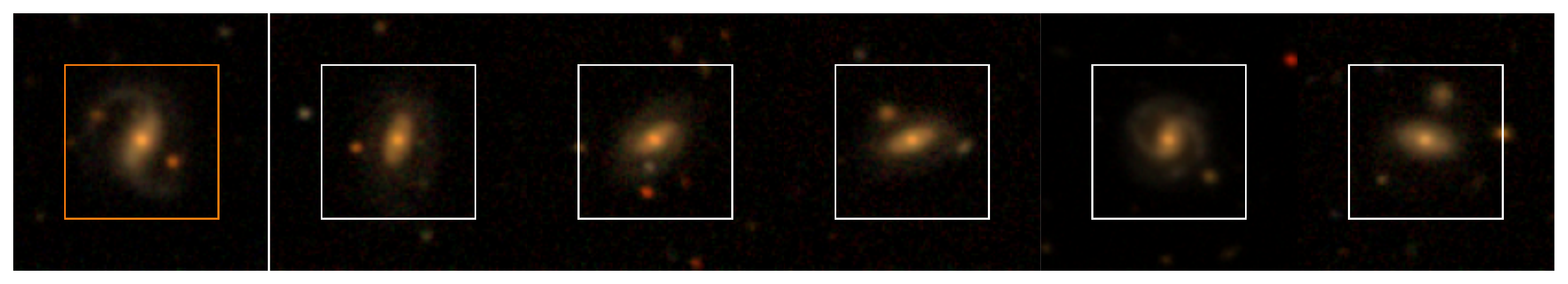}
    
    \includegraphics[width=0.4\textwidth, 
    trim={0.25cm 0.25cm 0.25cm 0.25cm}, clip]{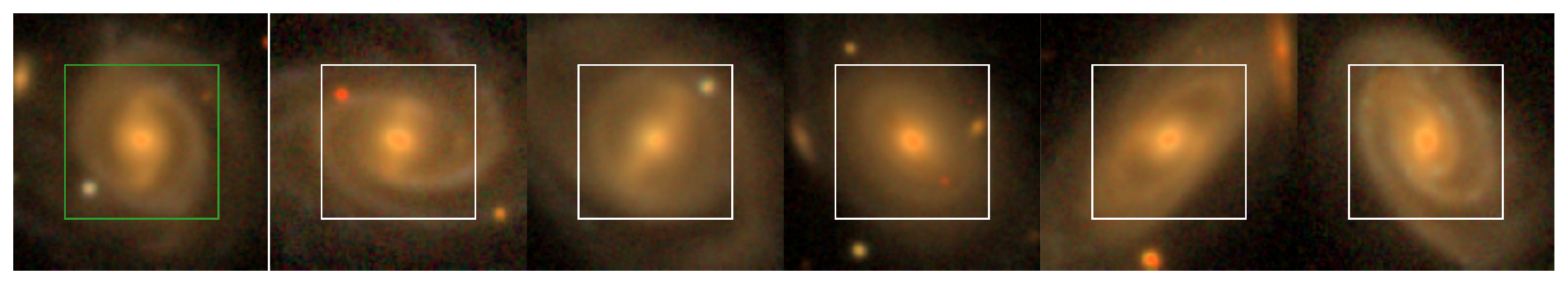}
    \includegraphics[width=0.4\textwidth, 
    trim={0.25cm 0.25cm 0.25cm 0.25cm}, clip]{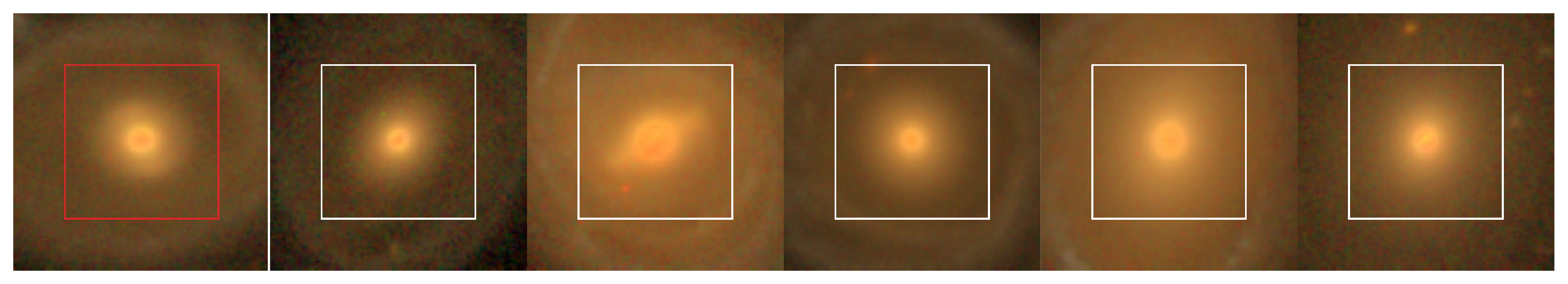}

    \includegraphics[width=0.4\textwidth, 
    trim={0.25cm 0.25cm 0.25cm 0.25cm}, clip]{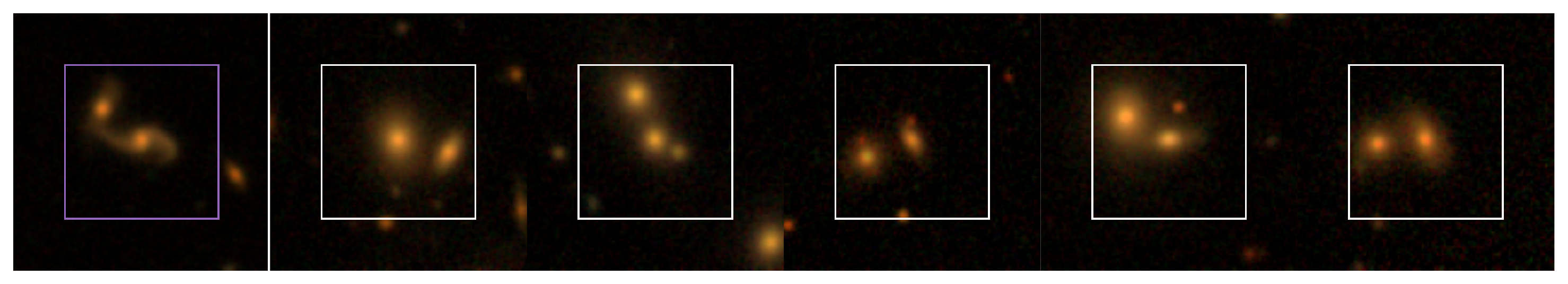}
    \includegraphics[width=0.4\textwidth, 
    trim={0.25cm 0.25cm 0.25cm 0.25cm}, clip]{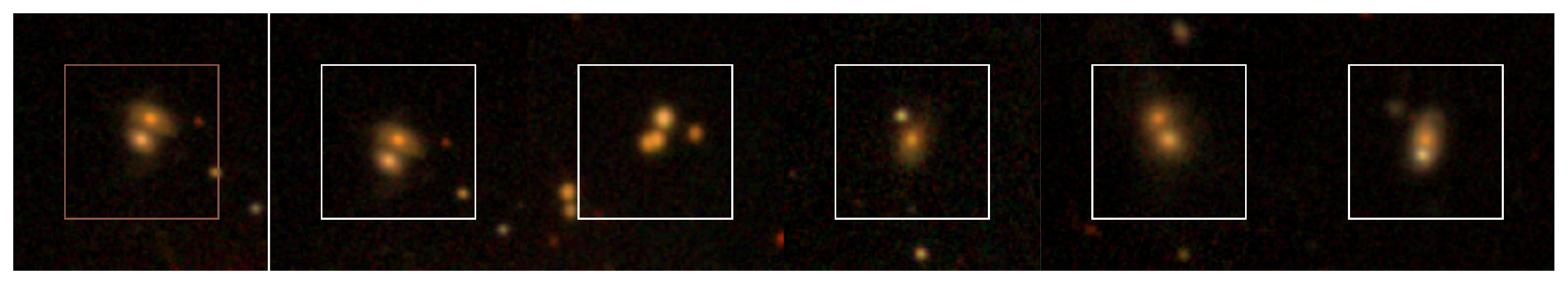}
    
    \includegraphics[width=0.4\textwidth, 
    trim={0.25cm 0.25cm 0.25cm 0.25cm}, clip]{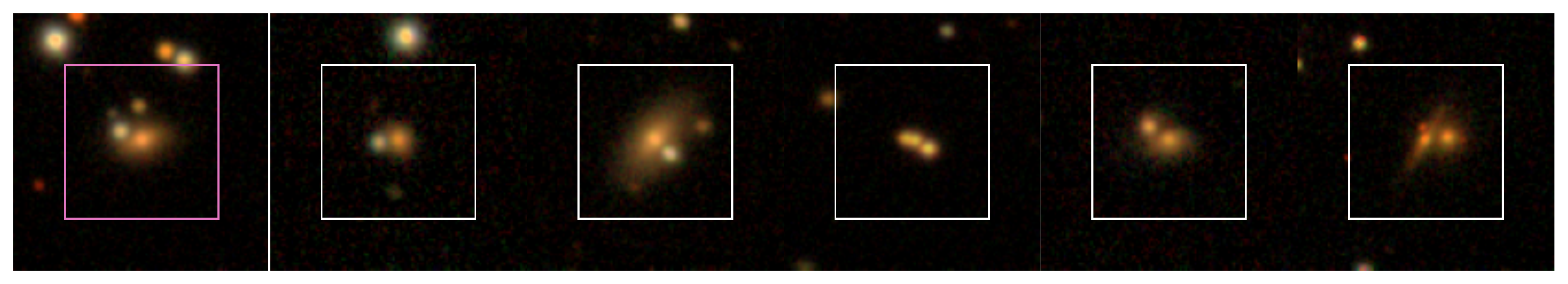}
    \includegraphics[width=0.4\textwidth, 
    trim={0.25cm 0.25cm 0.25cm 0.25cm}, clip]{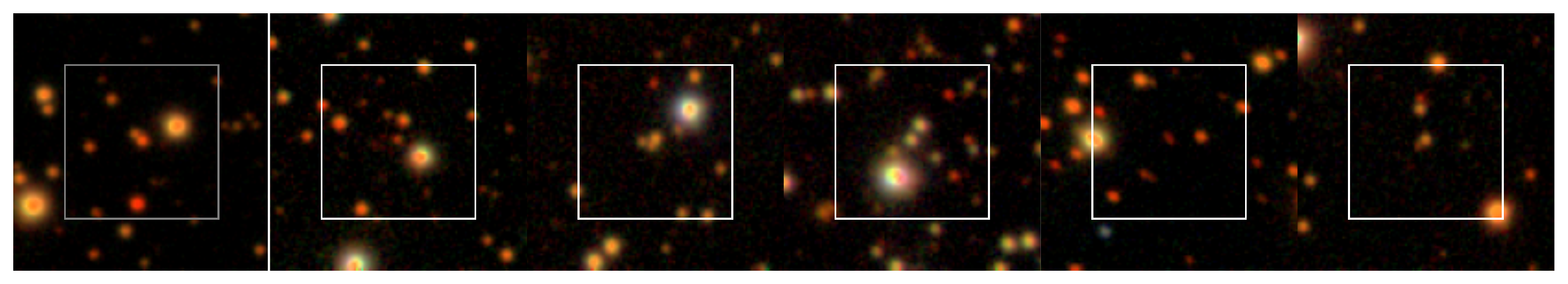}
    
    \includegraphics[width=0.4\textwidth, 
    trim={0.25cm 0.25cm 0.25cm 0.25cm}, clip]{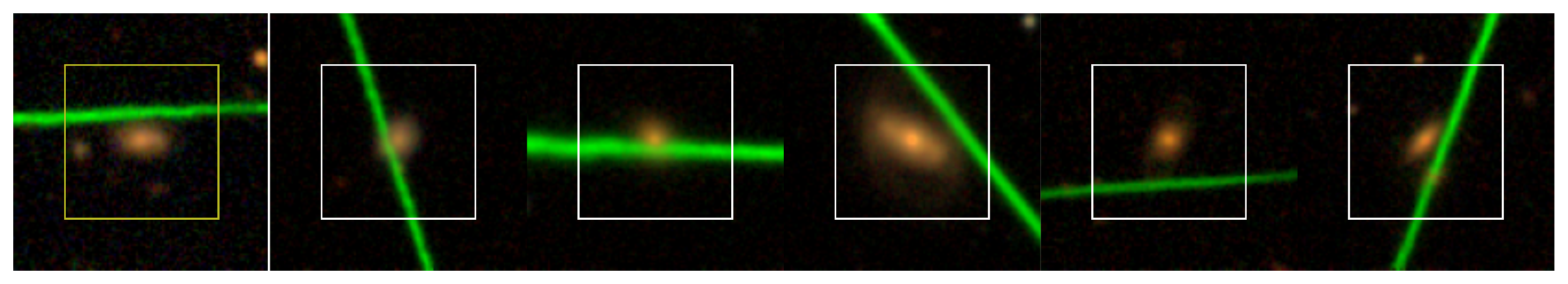}
    \includegraphics[width=0.4\textwidth, 
    trim={0.25cm 0.25cm 0.25cm 0.25cm}, clip]{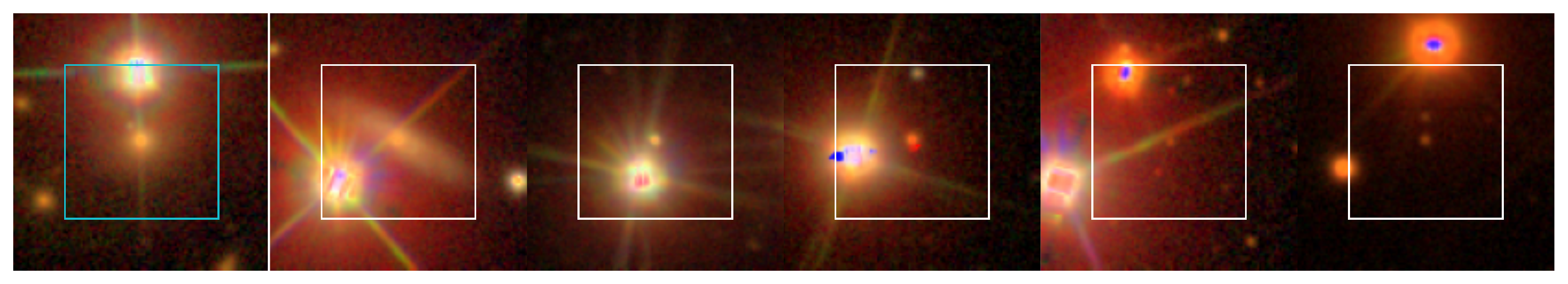}
    
\vspace{-7pt}
  \caption{Reference SDSS galaxies (leftmost panels) and the most similar galaxies (following 5 panels) identified through a self-supervised similarity search. Squares outline the 64$^2$ pixels that are ``seen'' by the network.}
  \label{fig:similarity}
\end{figure}

As previously mentioned, the self-supervised representations provide a venue to perform similarity searches. The contrastive framework that the encoder was trained under worked to encode images that are semantically similar to nearby points in the representation space. Therefore, given any desired query image, finding semantically similar images that may exist in the dataset corresponds to finding nearby data points in the 2048 dimensional representation space. This can be achieved by taking the cosine similarity (i.e. the normalized dot product) of the query vector and all other representations, and sorting by decreasing value. Alternative similarity measures to the cosine similarity can also be used, but were not studied here.

Figure~\ref{fig:similarity} demonstrates a similarity search performed on ten different examples of SDSS images. The images returned by this simple similarity search, in a completely unsupervised fashion, visually appear extremely similar, and are seemingly agnostic to rotations and jitter, as desired.

\subsection{Image-based UMAP}
\label{sec:umap_image}

\begin{figure}[t]
  \centering
    \includegraphics[width=0.8\textwidth, clip]{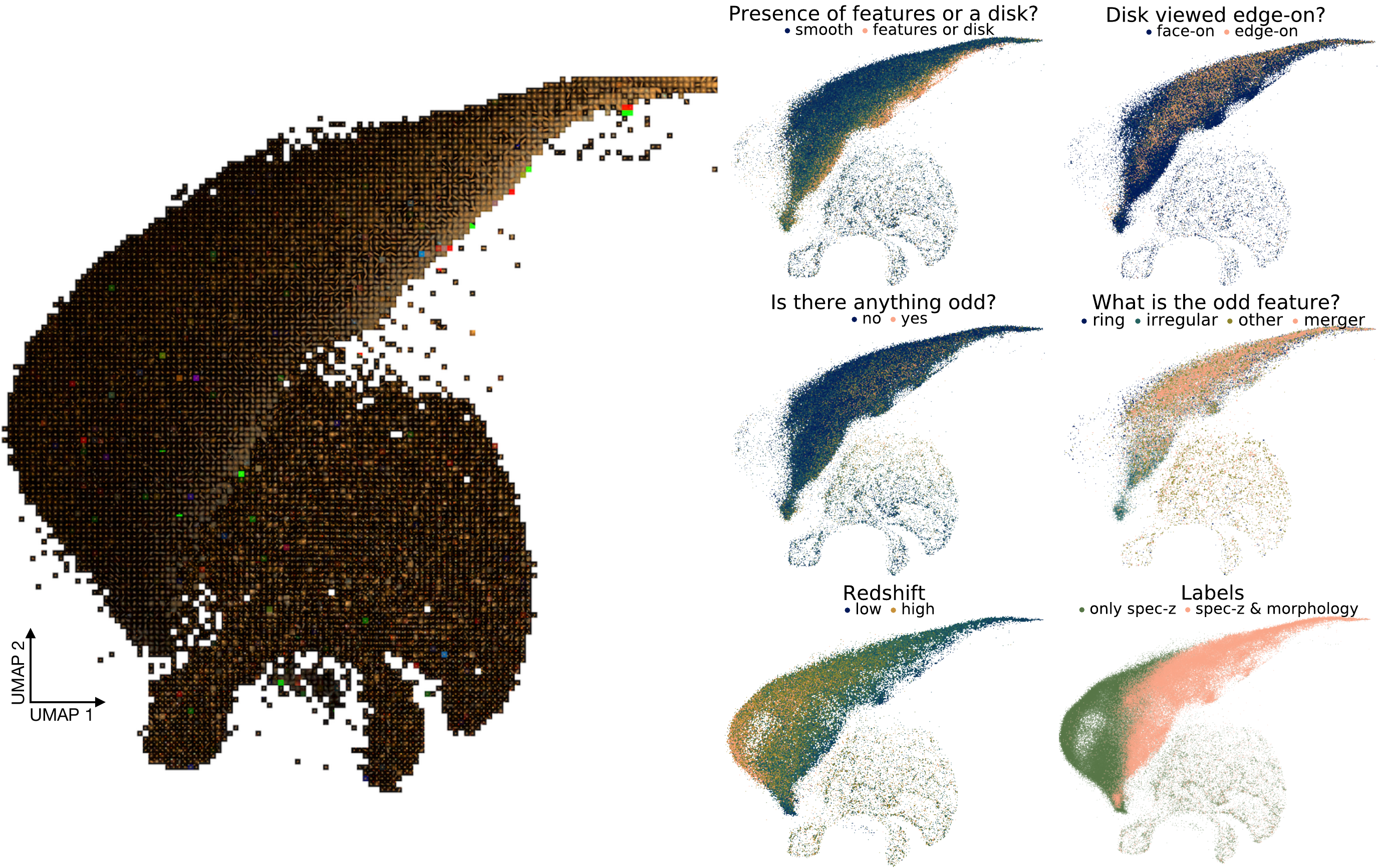}
  
 \caption{Visualizing the two dimensional UMAP projection of the image pixel space. The left panel shows randomly sampled representative images at each point in the space, while the right colors the space using answers to morphological classification questions from Galaxy Zoo 2, SDSS spectroscopic redshifts, or by labels.}
  \label{fig:umap_images}
\end{figure}    

Similar to the UMAP analysis on the self-supervised representations, UMAP can also be applied directly in pixel space by flattening the 5 band images into a vector. Due to computational difficulties with such large vectors we randomly sampled 5\% of the galaxies in the total dataset to determine the UMAP transformation, then used the transformation on the entire set of $\sim$1.3 million flattened images (labeled and unlabeled).

Figure~\ref{fig:umap_images} shows the equivalent of Figure~\ref{fig:umap}, but directly derived from the images. Here we only show galaxies that have labels. In the left panel we find that the images have been separated by the size and brightness of the galaxy, with bright and large galaxies residing in the top-right of the two dimensional space and dim small galaxies along the left. In the right panels we color each point by a subset of available labels and find that the first morphological classification - ``presence of features or a disk'' - shows a level of separation between the classes, while more detailed morphological features do not. Separation is also seen by redshift and label type, which are both strongly correlated with galaxy size and magnitude. 

While an unsupervised dimensionality reduction technique such as UMAP performed on the images can provide some level of clustering based on the semantic information within the images, it by definition does not attempt to map semantically similar images to similar points in the compressed space. To explicitly illustrate this undesirable outcome of an unsupervised clustering method we select a sample of galaxies and apply three sets of augmentations to each. The first rotates each image between 0 and 360 degrees in 45 degree increments, the second jitters each image from (-7,-7) to (7,7) pixels in 7 linearly spaced values and then crops the central 64$^2$ pixels, and the third adds Gaussian noise with a standard deviation ranging from 0 to 3 times the aggregate median absolute deviation measured over the dataset. Both the pixel space and representation space UMAP transforms were trained on 5\% of the dataset to facilitate a fair comparison. 

\begin{figure}[t]
  \centering
    \includegraphics[width=0.49\textwidth, clip]{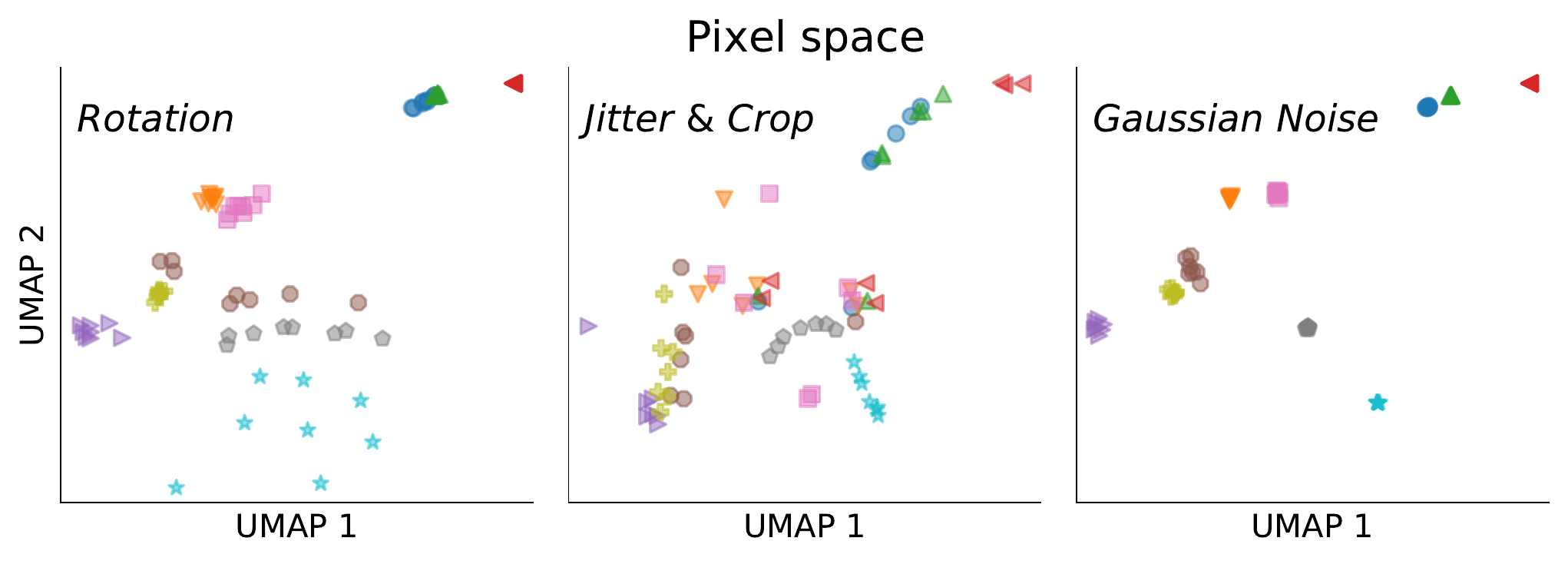}
    \includegraphics[width=0.49\textwidth, clip]{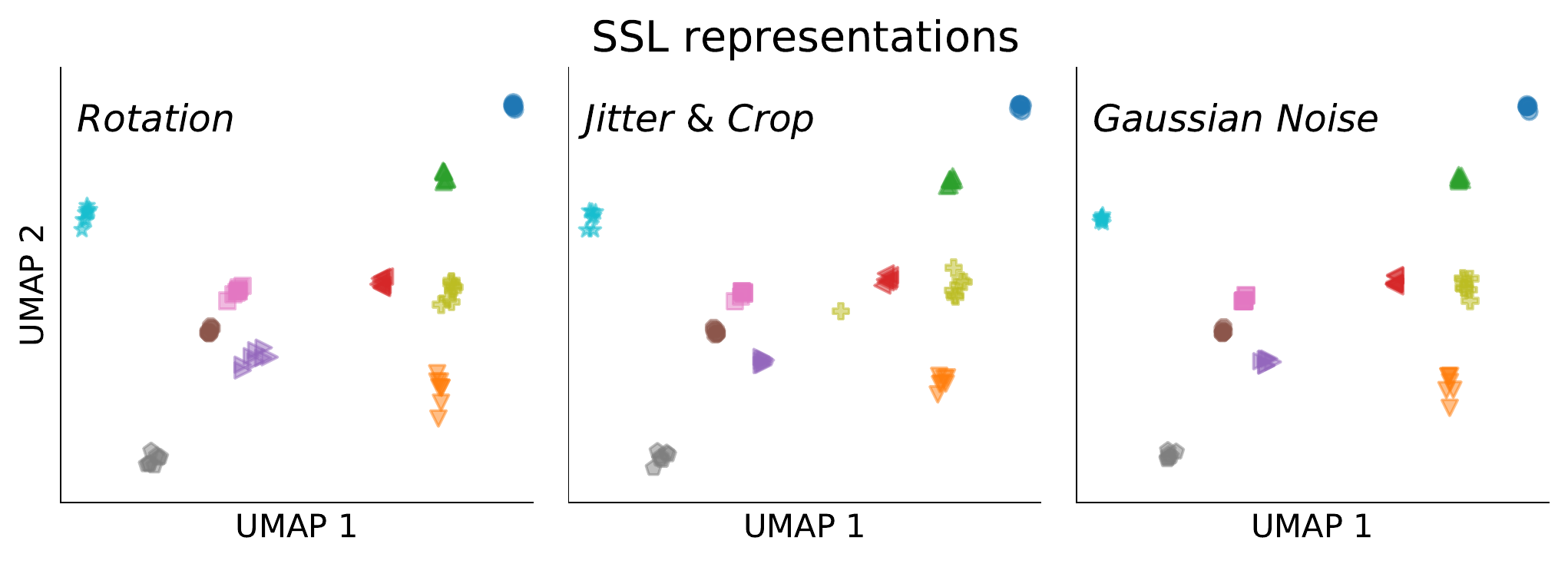}
    
 \caption{Stationarity of selected galaxies under data augmentations. Rotations were chosen uniformily between 0 and 360 deg, jitter from (-7,-7) to (7,7), and Gaussian noise with a standard deviation ranging from 0 to 3 times the aggregate median absolute deviation measured over the dataset. Colors match those of the left-most panels of Figure~\ref{fig:similarity}, as we used the same 10 galaxies.}
  \label{fig:augmentations_umap}
\end{figure}

Figure~\ref{fig:augmentations_umap} shows how the 10 galaxies used in the similarity search figure move through the UMAP plane under three different types of data augmentation. The marker color is consistent with the color of the square around the query image in Figure~\ref{fig:similarity}. The left three panels are created from the pixel space, while the right three panels are from the self-supervised representations. For the pixel space UMAP we see that the simple augmentations of rotation and jitter/crop cause the test galaxies move through a large fraction of the two dimensional plane, even though the semantic qualities of the galaxy remain unchanged. For the self-supervised representations we instead find that they are nearly invariant to the augmentations, which is expected due to the contrastive learning framework used to train the encoder. Only one augmented instance of a single galaxy shows any significant movement in the plane - the yellow plus symbol under the jitter/crop augmentation. This is the image shown at the bottom left of Figure~\ref{fig:similarity}, which is contaminated by a large green streak across the top third of the image. Under a large negative jitter, this streak no longer appears in the 64$^2$ pixel image fed to the encoder, so the representation changes. This demonstrates the advantage of contrastive self-supervised representation learning over common unsupervised clustering techniques.

\subsection{Additional Augmentations}

\begin{figure}[t]
  \centering
      \includegraphics[width=.24\textwidth, clip]{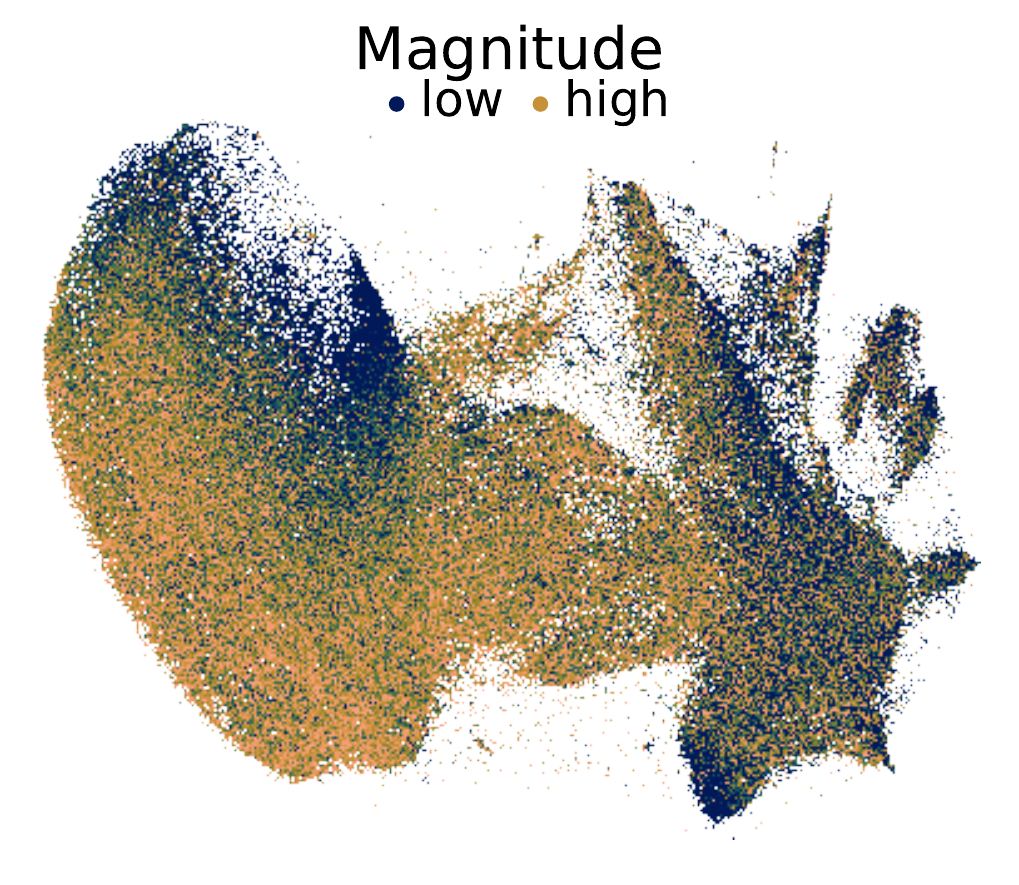}
    \includegraphics[width=.24\textwidth, clip]{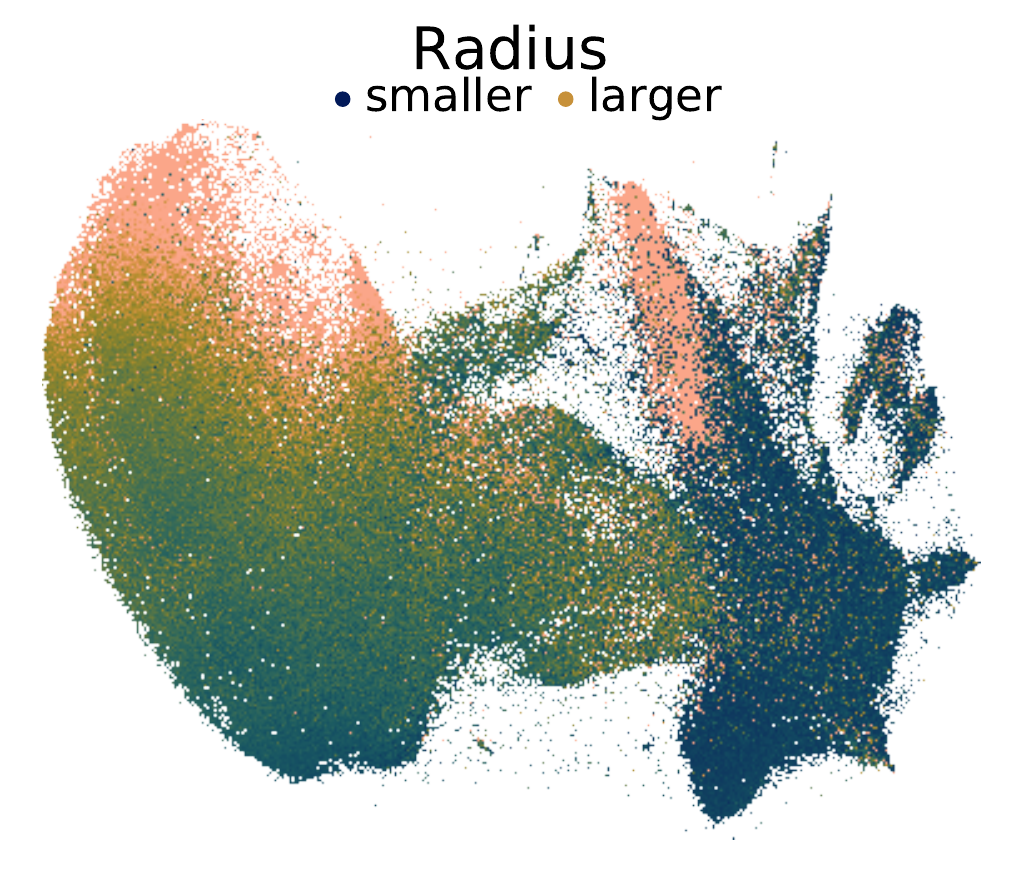}
       \includegraphics[width=.24\textwidth, clip]{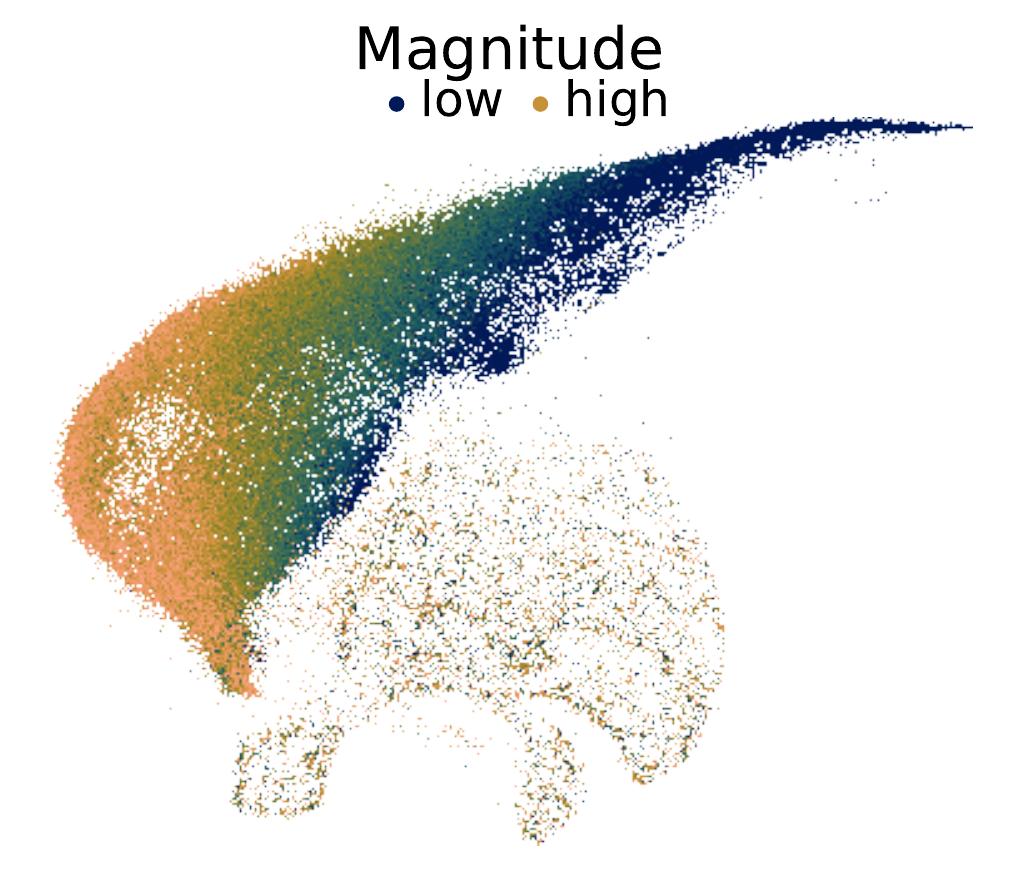}
    \includegraphics[width=.24\textwidth, clip]{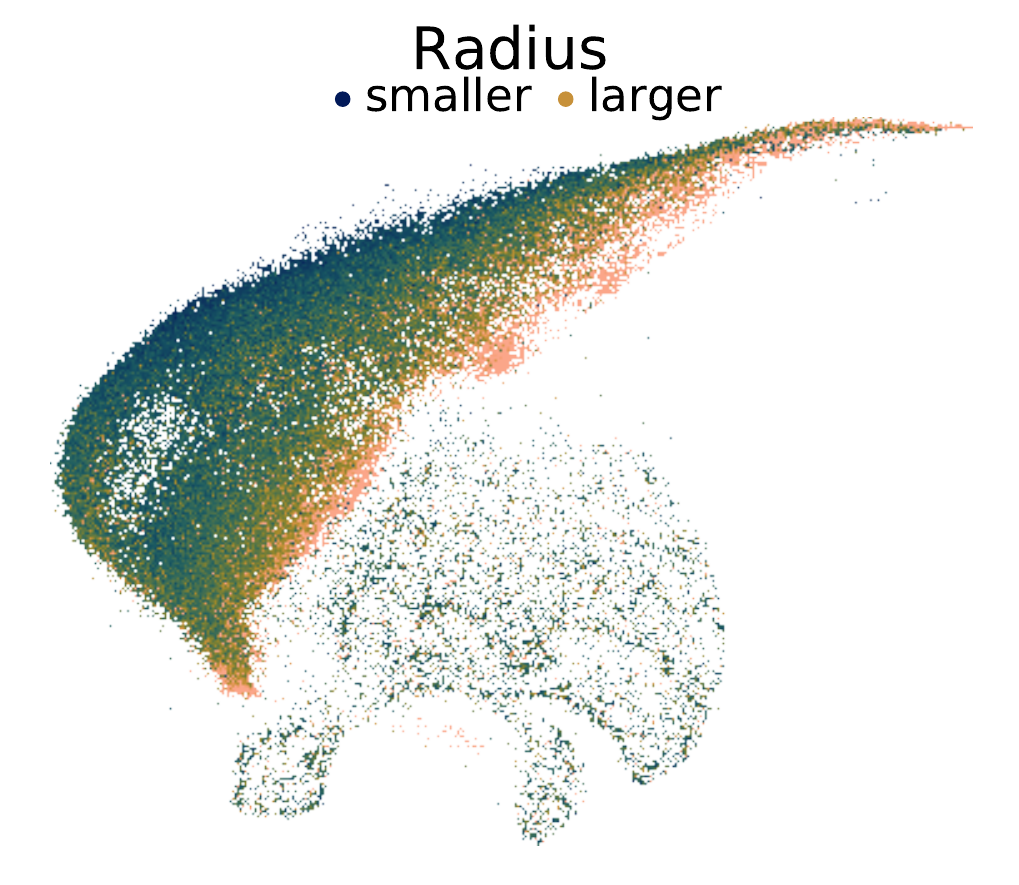}
     
 \caption{Representation (left) and image (right) UMAPs colored by r-band petrosian magnitude and radius.}
  \label{fig:umap_extras}
\end{figure}    

To remain as downstream task-agnostic as possible we only implemented the base set of augmentations described in Section~\ref{sec:method}. This has consequences on the information contained within the final self-supervised representations, and for specific targeted downstream applications additional augmentations may be useful. For example, morphological classifications should not depend on the size of a galaxy, and it is common to include a re-scaling augmentation when training a supervised classification network for this task. However, galaxy size is a useful indicator of redshift for nearby galaxies and is desirable in similarity search applications. Therefore task-specific augmentations beyond the base set can instead be added during the fine-tuning process.

Figure~\ref{fig:umap_extras} colors each point in the UMAP space by the r-band petrosian magnitude and radius to shows that the representations respond to both the size and magnitude of the galaxy. This is expected as we did not include a size scaling augmentation, and no normalization was performed on the images to remove magnitude information. When comparing this figure to the previous UMAP visualizations it is apparent that there are strong correlations between the radius, magnitude, redshift, presence of features or a disk, and the available labels. When pushing the photometric redshift prediction to higher redshifts in future works, where the size and magnitude of an object are not as good of indicators of redshift, a further study on the utility of size and magnitude augmentations for this task will be required.

\section{Galaxy Morphology Classification}\label{app:morphology}
\label{ap:galaxy_morphology}

In Section~\ref{sec:morphology} we displayed morphological predictions for the first Galaxy Zoo 2 (GZ2) questions. Here, we describe the training data and detail the training procedure for the linear classifier, denonstrate the results on two additional galaxy zoo questions, and put in further context the accuracy of the predictions with respect to previous methods.

\subsection{Training methodology}
Galaxy Zoo 2 \citep{GZ2} achieved 16 million morphological classifications of 304,122 galaxies drawn from SDSS with r $<$ 17 and petroR90\_r\footnote{petroR90\_r is the Petrosian radius which contains 90\% of the r-band flux} $>$ 3 arcsec, collected via a web-based interface. For each galaxy users were shown a 424$\times$424 pixel image scaled to 0.02petroR90\_r arcseconds per pixel, and were led down a multi-step decision tree to answer increasingly detailed questions about the visual appearance of the galaxy. See Figure 2 of \citep{GZ2_debiased} for a clear illustration of the decision tree. For example, users were first asked ``Is the galaxy simply smooth and rounded, with no sign of a disk?'', and had the option of selecting responses ``smooth'', ``features or disk'', or ``star or artifact''. Their selection of a response is referred to as their {\textit{vote}}. Based on their response to a question they are lead down different branches of the decision tree - if ``smooth'' was selected, they are next asked ``how rounded is it'', but if ``features or disk'' was selected they will be asked a series of questions about spiral arms, bulge shapes, and a variety of other morphological features.

There are a total of 11 {\textit{classification tasks}} (questions) that have the potential of being asked in the GZ2 decision tree, and the number of possible responses for each classification ranges from 2 to 7. We focus on binary tasks, with the exception of the first question where the response ``star or artifact'' occurs so infrequently (0.08\% of responses select this as the majority) that it can be effectively neglected.  All tasks beyond the first one depend on responses to previous tasks in the decision tree. For example, ``could this be a disk viewed edge on?'' is only asked if the user responded ``features or disk'' on the first task. Thus, a response for ``edge-on'' over ``face-on'' is not a binary classification of the total galaxy population, but only a sub-classification of galaxies that were already considered to have features or a disk.

The nature of data collection (non-expert labelled), coupled with the uncertain class boundaries for galaxies with faint features, result in individual GZ2 users voting for different answers, and therefore uncertain morphological classifications are given for each galaxy. We consider the ``consensus weighted vote fractions'' - the fraction of users who voted for an answer\footnote{consensus weighted fractions are slightly different than the true vote fraction, they are the result of re-weighting users votes based on their overall consensus with others who looked at the same image} -  as the true probability of a galaxy belonging to one class over the other, and we predict an equivalent class probability between (0, 1). Rather than a binary prediction, the returned probability represents the uncertainty of the morphology of the galaxy as seen in an SDSS image, whether this uncertainty stems from faint features, mislabelling, or imaging artifacts. Other works use the ``de-biased estimate'', which estimates how the galaxy would have been classified if viewed at $z=0.03$ \citep{GZ2_debiased}. By using the consensus weighted fractions we estimate what the image actually shows, not the ``true'' morphology, and debiasing can be performed after prediction. 

We use the GZ2 main sample with spectroscopic redshifts which includes morphological classifications for 243,500 galaxies. The main sample without photometric redshifts of 42,462 galaxies, and the stand-alone Stripe 82 catalogue of 17,787 galaxies, were not included. We cross match the GZ2 table\footnote{Galaxy Zoo data is located at \url{https://data.galaxyzoo.org/}} with our SDSS database and search for pairs whose equatorial coordinates overlap within within 5 pixels (1.98 arcsec). This returns a final sample of 183,929 galaxies for which we have both \textit{ugriz} images and crowd-sourced morphological classifications. This sample also contains redshift information which was used separately for photo-z prediction. For each question we required a minimum of 5 votes for each galaxy, which results in questions that are less frequently answered having smaller number of labelled samples than the full 183,929.

\subsection{Network \& training}

We predict the vote weighted user response using three different classifiers. The first is a CNN trained from scratch in a supervised setting with the same ResNet50 architecture of the encoder, the second is a linear classifier {\textit{directly on the self-supervised representations}}, and for the third we fine-tune the self-supervised encoder for a few epochs. Classifcation is performed by the addition of a fully connected layer on the 2048 dimensional output of the ResNet50. This maps the 2048 dimensional representation to 1 dimension with 2048 trainable weights and one bias parameter, followed by a softmax function to ensure the predicted probability is within the range (0, 1). For each GZ2 question networks are trained separately using the subset of galaxies that have at least five total votes for that question. Thus, the predicted probability should be interpreted as follows: regardless of the vote fraction of responses to previous GZ2 questions, what is the consensus vote {\textit{of the users that ended up at this question on the decision tree}}. If only 10\% of users selected ``features or a disk'' for the first question, the galaxy most likely is smooth and does not have a disk. Yet we still use the vote fraction of that 10\% of users that were then subsequently asked ``could this be a disk viewed edge-on?'' when training a classifier for the edge-on question. Classifiers can then be used in conjunction after training.

Training was conducted in PyTorch~\citep{pytorch} through a binary cross entropy loss on the soft labels (consensus weighted vote fractions). A random sample of 20\% of the data was set aside for testing and was not used for training any of the classifiers, and 10\% was set aside for validation. For the supervised and fine-tuned networks we augmented images at each training epoch with jitter/crop and random rotations. Many questions have a high degree of class imbalance which reflects the occurrence of galaxy morphologies in the nearby universe. We found that class-balanced class weights (each instance of the class weighted by the overall occurrence fraction of that class) did not improve the classification performance. We weighted each instance by the total number of votes it received, although this resulted in negligible performance differences. 

The supervised network was trained on 8 NVIDIA Tesla V100 GPUs for 100 epochs with a batch size of 128 using the SGD optimizer with a learning rate of 0.01, which we reduced by a factor of 10 at 60 and 90 epochs. The fine tuned-network was trained similarly, but with the pre-trained weights having a learning rate 10x smaller than the linear classifier layer. Optimization for the linear layer directly on the representations was performed using Limited-memory BFGS (LBFGS) and  a learning rate of 0.05, which was decreased by a factor of 10 at 10 and 25 epochs. For all networks we used the epoch that produced the highest accuracy on ``high quality'' labels in the validation set. Supervised ResNets and fine-tuning required between a few minutes and a few hours of training time on the 8 GPUs, depending on the number of training samples. For each linear classifier on the representations training generally concluded within a few epochs and $\sim 0.5-60$ seconds of compute time on a GPU. Note that this does not take into account the compute time required to learn the self-supervised representations, which this is shared between all downstream tasks, and does not need to be undertaken for each one separately. 

As evidenced by Figure~\ref{fig:umap}, the self-supervised representations have achieved a high degree of separation between numerous types of galaxy morphologies. Likewise, we found that classification became a straightforward task when using any subset of the labels. Unlike DS+18, we found no need to separate uncertain labels from high probability ones. Limiting our training set to only `high quality' (HQ) labels (those with $P<0.2$ or $P>0.8$), resulted in nearly equivalent performance on HQ labels in the test set, but significantly decreased the performance on uncertain labels (those with $0.2 < P < 0.8$). We also found no need to impose higher minimum cuts on the number of votes (10) for some questions, as W+20 found was needed to improve classification performance from random initialization. 

\subsection{Performance metrics \& additional results}

Figure~\ref{fig:morphology} demonstrated the quality of the morphological predictions for the first GZ2 question, and here we include the results on the second and third questions. To measure the performance of a binary classifier, we quantify the accuracy (Acc), precision (Prec), recall or true positive rate (Rec/TPR), false positive rate (FPR), area under the receiver operator characteristic curve (AUC), and the outlier fraction $\eta$:
\begin{align*}
    \text{Acc} &= \frac{\text{TP} + \text{TN}}{(\text{TP}+\text{FP}) + (\text{TN} + \text{FN})} \\
    \text{Prec} &= \frac{\text{TP}}{\text{TN}+\text{FP}} \\
    \text{Rec/TPR} &= \frac{\text{TP}}{\text{TP}+\text{FN}} \\
    \text{FPR} &= \frac{\text{FP}}{\text{FP}+\text{TN}} \\
    \text{AUC} &= \int \text{TPR} d(\text{FPR}) \\
    \text{$\eta$} &= \frac{[(P_{true} \leq 0.2) \cap (P_{predict} \geq 0.8)] + [(P_{true} \geq 0.8) \cap (P_{predict} \leq 0.2)]}{N_{HQ}} \times 100, 
\end{align*}
Where TP and TN are the number of true positives and true negatives, and FP and FN are the number of false positives and false negatives, respectively, P$_{true}$ and P$_{predict}$ are the true and predicted labels, and N$_{HQ}$ is the total number of labels considered high quality. We use a straight probability cut of 0.5 for both the true labels and predictions (P $<$ 0.5 belongs to class 0 and  P $\geq$ 0.5 belongs to class 1), although quoted results can be slightly improved by allowing the probability cut on the predicted labels to vary.  

\begin{figure*}[t]
\centering
\includegraphics[trim={0cm 0 0 0},clip, width=0.8\textwidth]{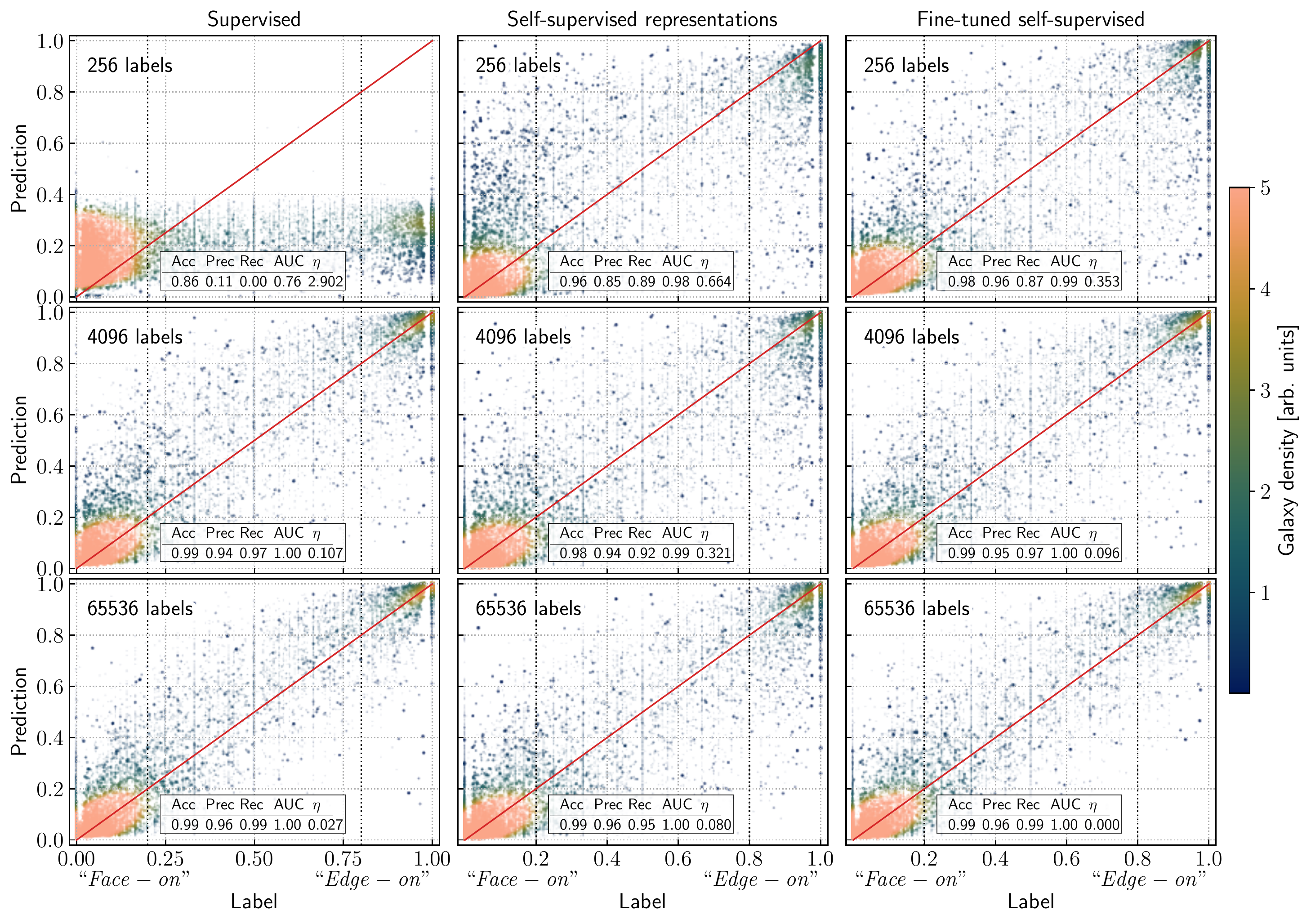}
\caption{Predicted labels compared to crowd-sourced answers for the first GZ2 question: ``Could this be a disk viewed edge-on?''. The size and opacity of the points is proportional to the number of crowd-sourced labels they received.}
\label{fig:morphology_q2}
\end{figure*}

\begin{figure*}
\centering
\includegraphics[trim={0cm 0 0 0},clip, width=0.8\textwidth]{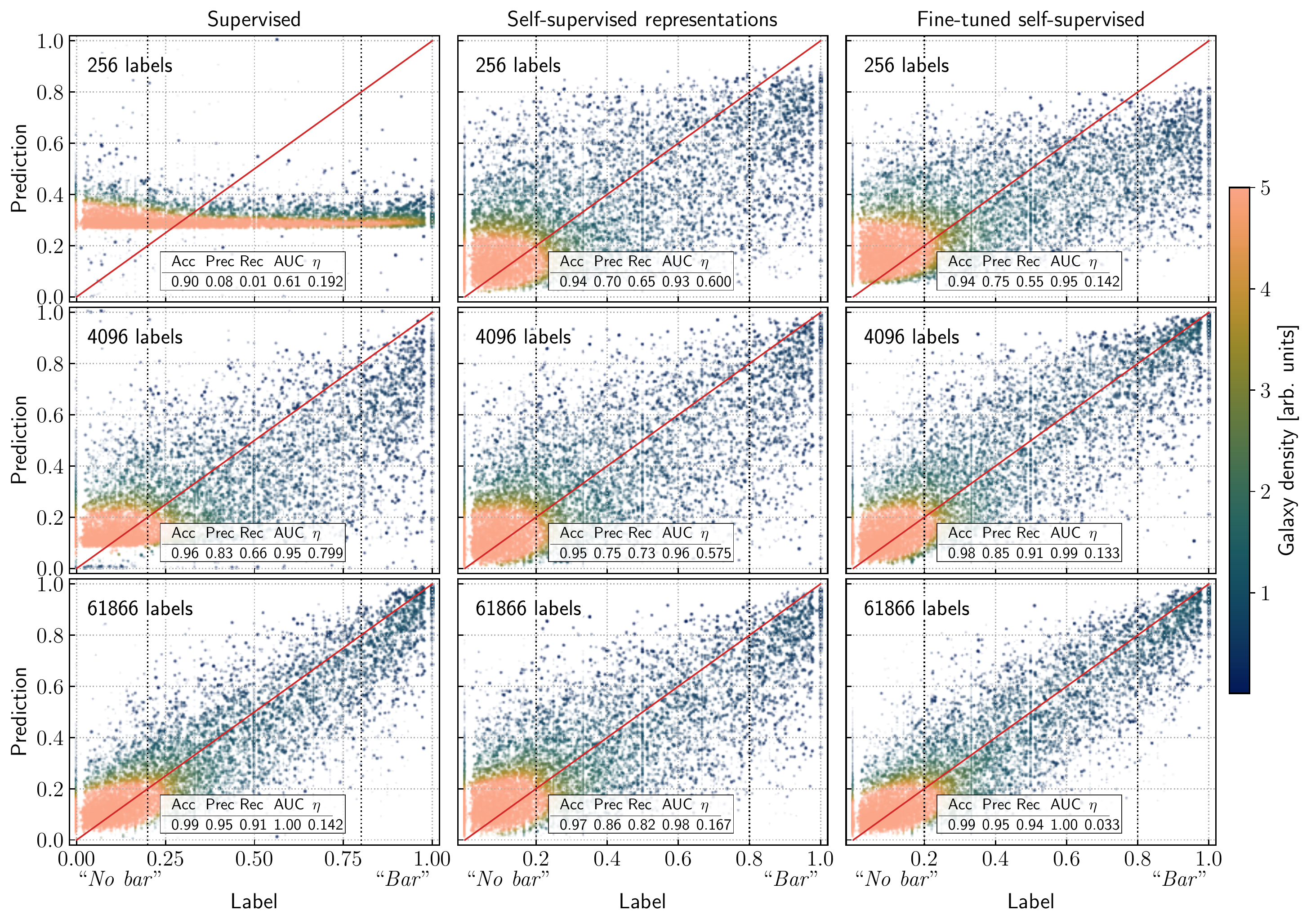}
\caption{Predicted labels compared to crowd-sourced answers for the third GZ2 question: ``Is there a sign of a bar feature through the centre of the galaxy?''. The size and opacity of the points is proportional to the number of crowd-sourced labels they received.}
\label{fig:morphology_q3}
\end{figure*}

Figures~\ref{fig:morphology_q2} and ~\ref{fig:morphology_q3} show the classification results in the same style as Figure~\ref{fig:morphology}, but for the second and third GZ questions, respectively. For the second question we find that 256 labels are sufficient to train the linear classifier on the representations and the fine tuned network to a high degree of accuracy, while the supervised network does not converge. This question, ``edge-on or face-on'' is likely the `easiest' of all GZ2 questions with obvious differences between the classes, and we find that 4096 labels is sufficient to train all three networks to a high accuracy. The third question, `bar or no-bar', is one of the most difficult from GZ2  due to uncertainty in human labelling and ambiguity for all but the most-obvious bars. It also suffers from the largest class imbalance of the three, and 256 training samples only contains 17 high quality positive samples, and 38 with label probability ranging from 0.5 to 0.8. Nonetheless, accurate results are still achieved with limited training samples. In all cases the fine-tuned self supervised model outperforms its supervised counterpart, and with a limited number of labels a simple linear layer on the representations outperforms a fully supervised network. 

We note that an exact quantitative comparison of our performance metrics to DS+18 and W+20, or to other automated morphological classification works, is not possible due to a lack of consistency in data sets. The samples used from the full set of GZ2 answers are not consistent: images can be from different SDSS data releases and have inconsistent pixel sizes or number of pixels, they use different observation bands (we use 5 here, while most works use 3) and different image normalization/re-sizing is applied, and different GZ2 vote fraction definitions are used as the ``ground truth'' morphological classification. Most importantly, the class imbalance used in training/testing is significant (generally lower in our work as we use all samples), and more imbalanced datasets will more easily show up as higher metric scores. Nevertheless, the results achieved here are extremely promising, and show that the use self-supervised representations allows us to push beyond the limits of supervised learning for automated morphological classifications when limited by the number of available labels. 

We have performed a number of tests to ensure that the networks are not basing their classification on secondary characteristics. The results shown are from un-normalized versions of the images, which include color information based on the relative brightness in each channel. We performed the same suite of classification exercises on images where we first normalized each channel of each image by the maximum pixel value, and found negligible differences to the results shown here. As seen in both UMAP figures, at least the first GZ2 question `features or disk'' has labels that are correlated with both galaxy size and magnitude. To test if a simple model using mainly these pieces of information can achieve high classification results; we trained a UMAP on images from the test set to reduce to 2048 dimension representations in an unsupervised fashion. Using these UMAP representations as inputs for a linear classifier, equivalent to how we trained out linear classifier on the self-supervised representations, we found that performance never increased above initialization. These tests confirm that a simple combination of size/brightness/color is not enough to achieve accurate predictions for these classifications.

Misclassified galaxies, specifically those that are large outliers, have three main failure modes which can be addressed in future work to further improve results beyond those achieved here. Firstly, our images span 25.3 arcseconds, which is not large enough to cover the entire angular extent of very nearby and large galaxies. In contrast, GZ2 participants were shown images scaled to ensure the entire galaxy was always visible, as do other automated classifications. Classifying these limited number of these galaxies is easily achievable by human means, so targeting these few samples is not of top priority. But, for the specific downstream task of classifying galaxy morphologies an additional augmentation which scales the angular extent of galaxies may prove beneficial. Second, a number of misclassified galaxies have imaging artifacts, which fine tuning the network would likely help improve the classification. A similarity search on the self-supervised representations provides a very valuable tool to isolate these artifacts. Finally, some `misclassified' galaxies are the result of label uncertainty, which is especially rampant when the true label is outside of the high quality range.


\section{Data augmentation ablation study}\label{app:ablation}
In Section \ref{sec:method} we have listed a basic set of augmentations which make intuitive sense for use in a contrastive learning framework, as each augmentation is associated with some source of observational variability within the images that we want the learned representations to be invariant to. A relevant task in developing the self-supervised model is deciding which combinations of these augmentations are most effective in producing high-quality, semantically useful image representations. This can be answered in a number of ways, but is traditionally done in computer vision by taking a sample downstream task, like image classification, and training a linear classifier on top of the learned representations to perform it. Doing so evaluates the quality of the learned representations by measuring, e.g., how easily different classes are linearly separated in the representation space.

Such an approach is straightforward for the task of morphology classification, but is slightly ill-conceived for something more challenging like photo-z estimation, since the ``classes’’ output by the network represent consecutive redshift bins which should not necessarily be linearly separable. Instead, we evaluate our representations by fine-tuning them for the photo-z estimation task and using the $\sigma_{\rm MAD}$ of predictions on test data as our quality metric. To ensure this metric is more closely tied to the representation quality rather than the supervised fine-tuning process, we only consider the performance of models which are fine-tuned on 10\%, 20\%, and 30\% of the labeled data.

\begin{figure}
    \centering
    \includegraphics[width=0.4\linewidth]{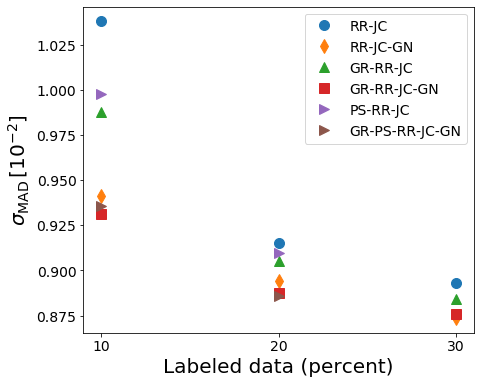}
    \caption{The $\sigma_{\rm MAD}$ of photo-z predictions from models fine-tuned on various fractions of the labeled data, showing how different combinations of augmentations affect downstream performance. Augmentations shown are random rotation (RR), jitter crop (JC), Gaussian noise (GN), galactic reddening (GR), and point spread function (PS).}
    \label{fig:ablation}
\end{figure}

The results of our evaluation are shown in Figure \ref{fig:ablation}. We find that on our dataset, the Gaussian noise augmentation seems to be the strongest, but that the best performance is achieved when we use all augmentations except the PSF smoothing. An important note here is that this finding depends both on our dataset as well as the way we have implemented each augmentation. Using contrastive learning with other surveys would require different implementations and possibly different augmentations, and could produce different ``hierarchies’’ of augmentation strengths. In general, careful thought needs to be put into the data augmentations and which transformations one wants the representations to be invariant against \citep{Xiao2020}.

\section{Additional architecture, hyperparameter, \& training details}
\label{app:architecture}
\subsection{Self-supervised framework}\label{app:encoder}
\textbf{Encoder:} We follow~\citep{he2020momentum, chen2020simple} in using a ResNet50 architecture~\citep{he2016deep} as our encoder. Specifically, we use the implementation of the TorchVision library ({\tt torchvision.models.resnet50}) which is a part of the PyTorch project~\citep{pytorch}. The standard ResNet50, however, is designed to work on wider images than our $64\times64$ ones. To maintain reasonably wide representations by the end of the 50 layers, we change the first {\tt Conv2d} layer to have {\tt stride=1} instead of the default {\tt stride=2}, we also remove the first {\tt MaxPool2d} layer. This gives us $4\times$ wider activations throughout the network than what we would get with the defaults of ResNet50. The output of the {\tt AdaptiveAvgPool2d} layer of the network is our representation, a $2048$ dimensional vector. We also change the number of input channels to 5 to work with our 5 $ugriz$ passbands data.

Following~\citep{chen2020simple,chen2020improved}, we don't use the representation ${\bf {z}}$ directly in the loss in Eq.~\ref{eq:loss}, instead, we use a two layer MLP projection head which maps the representations to a space where the contrastive loss is applied. This has been shown to improve the learned representations. The output of the projection head is a $128$ dimensional vector. The head is discarded after the self-supervised training process is completed.

\textbf{Momentum encoder:} 
In contrastive learning setups, in order to make the task of identifying positive examples non-trivial, it is crucial to have a large set of negative examples. For this we use the momentum encoder idea from~\citep{he2020momentum}; we maintain a queue of size $62$k representations ($\sim 5\%$ of the training dataset size) that is continuously being updated during the training process. The representations in the queue are encoded using a momentum encoder; a second encoder with same architecture whose weights are an exponentially moving average of the main encoder weights. The parameters $\theta_k$ of the momentum encoder network are updated using the encoder parameters $\theta_q$ with momentum parameter $m$ via

\begin{equation}
    \theta_k \leftarrow m\theta_{k}+(1-m)\theta_q. 
\end{equation}

The momentum hyper-parameter is set to $m=0.999$. The momentum encoder helps maintain some consistency between the representations in the negative examples queue during the training. We use temperature parameter $\tau=0.1$ in the contrastive loss.

During self-supervised pre-training, we use stochastic gradient descent with a cosine learning rate schedule, having an initial learning rate of 0.03. We pre-trained our network for 12 hours using 8 NVIDIA V100 GPUs on $\sim$1.3M images to complete $\sim$50 epochs and this proved to be good enough for learning useful features used in this study. We have used \texttt{DistributedDataParallel} of PyTorch to leverage distributed training.

\subsection{Photometric redshift estimation networks}
We closely follow the setup of~\cite{pasquet2019photometric}, whose CNN is trained as a classifier over a discrete set of 180 redshift bins of size $\delta z  = 2.2 \times 10^{-3}$ spanning $0\leq z \leq 0.4$, where the photo-z estimate $z_p$ is computed as the expectation $\mathbb{E}(z)$ over the probabilities predicted in each bin. We train models from scratch to establish a baseline with the ResNet50 architecture (with the same modifications made for the self-supervised pre-training encoder, see~\ref{app:encoder}).  During training, images are de-reddened by using the tabulated $E(B-V)$ value with the photometric calibration in \cite{schlafly2011measuring}, then augmented with random rotations, and random jitter \& crop. Only de-reddening and cropping is applied at testing or evaluation.

For the fine-tuned networks, we take the pre-trained encoder and replace the projection head with a single MLP layer. Here, parameters up to \texttt{AdaptiveAvgPool2d} are initialized with the pre-trained weights and the MLP layer has random initialization. The whole network is then trained on labels, with the pre-trained weights having a learning rate of $0.0001$, and the MLP classifier layer having a learning rate of $0.001$. We train for 100 epochs and reduce the learning rate by a factor of 10 at 60, 90-th epochs.  In all cases we have a batch size of 256.

\end{document}